\documentclass[aip,jcp,amsmath,amssymb,reprint,nofootinbib]{revtex4-2}
\usepackage[utf8]{inputenc}
\usepackage{amsmath}
\usepackage{amsfonts}
\usepackage{amssymb}
\usepackage{graphicx}
\usepackage{upgreek}
\usepackage{hyperref}
\usepackage{braket}
\usepackage{xr}
\usepackage{multirow}
\usepackage{siunitx}
\usepackage{lipsum}
\usepackage{color}
\newcommand{\etal}{\emph{et al.}}
\newcommand{\kB}{k_{\mbox{\tiny B}}}
\newcommand{\kBT}{k_{\mbox{\tiny B}}\hspace*{-0.33ex}T}

\renewcommand{\vec}[1]{\mathbf{#1}}
\newcommand{\m}[1]{\mathbf{#1}}
\newcommand{\re}{\mathrm{e}}
\graphicspath{{./fig/}}

\begin{document}
\small
\title{Coupled dynamics in binary mixtures of colloidal Yukawa systems}
\author{Daniel Weidig}
\affiliation{Institut f\"ur Chemie, Universit\"at Rostock, 18051 Rostock, Germany}
\author{Joachim Wagner}
\email{joachim.wagner@uni-rostock.de}
\affiliation{Institut f\"ur Chemie, Universit\"at Rostock, 18051 Rostock, Germany}

\date{\today}
\begin{abstract}
The dynamical behavior of binary mixtures consisting of highly charged colloidal particles is studied by means of Brownian dynamics simulations.
We investigate differently sized, but identically charged particles with nearly identical interactions between all species in highly dilute suspensions.
Different short-time self-diffusion coefficients induce, mediated by electrostatic interactions, a coupling of both self and collective
dynamics of differently sized particles: The long-time self-diffusion coefficients of a larger species are increased by the presence of a more mobile,
smaller species and vice versa. Similar coupling effects are observed in collective dynamics where in addition to the time constant of intermediate 
scattering function's initial decay its functional form, quantified by exponents of a stretched exponential decay, are influenced by the presence of a differently
sized species. We provide a systematic analysis of coupling effects in dependence on the ratio of sizes, number densities, and the strength of electrostatic 
interactions.
\end{abstract}

\maketitle

\section{Introduction}

Dynamical processes in strongly interacting systems and their relation to the time-averaged structure are since decades in the focus of scientific interest. Understanding of freezing processes, where correlations covering  orders of magnitude in time are crucial, is still a challenge both from the viewpoint of experiment and theory. With evolving photon correlation spectroscopy \cite{Berne:2000} as a quasielastic scattering method accessing timescales from microseconds to hours, colloidal suspensions gained interest as model systems on, compared to atomic or molecular systems, enlarged scales of lengths and times: The availability of highly defined particles with tailored interactions establishes a fruitful combination of experiment and theory to investigate structure-dynamics relations in interacting systems self-organizing to analogues of imperfect gases, liquids, glasses and even colloidal crystals.\cite{Pusey:1991,Tata:2001} Colloidal suspensions have become relevant model systems for an experimental test of mode coupling theory describing slow dynamic processes such as, e.g., the glass transition. \cite{szamel:1991,goetze:1992,vanmegen:1991,vanmegen:1991_1,DiazMaier:2024-2}

Opposite to model hard-sphere colloids interacting via their mutual excluded volume, highly charged colloidal particles interact via the long-range electrostatic repulsion of equally charged macroions. Hence, these 
systems form at volume fractions as low as $\varphi=10^{-3}$ colloidal crystals. \cite{Robbins:1988} According to DLVO theory,\cite{Verwey:1948} charged colloids interact via a screened Coulomb- or Yukawa-potential
while the mutual excluded volume is not relevant in dilute systems.

Binary mixtures of charged colloidal particles have previously been studied in two dimensions.\cite{Assoud:2008, Ott:2009, Kalman:2013, Ott:2014, Pankaj:2016} In three dimensions, studies are so far limited 
to hard-sphere systems \cite{Lynch:2008,Narumi:2011,Nguyen:2014,Yeo:2015,Sentjabrskaja:2019,medina:1988,Acuna:2000} while only few works investigate mixtures of charged particles.\cite{Krause:1992,Arrieta:1987,Vazquez:2003,Schoell:2005,Hopkins:2008,Kalman:2011} The vast majority of existing studies focuses on the self-dynamics of identical particles employing both, theoretical\cite{mendez:1999,Khrapak:2012,Khrapak:2018,Ohta:2000,Wang:2020,Sanz:2010,dzhumagulova:2013} and experimental\cite{vanBladeren:1992,wagner:2001_1,lellig:2004} methods. Collective dynamics are up to now only rarely analyzed.\cite{Rahman:1964,Gaylor:1980,Yeomans:2003,chavez:2006,Leshansky:2008,Banchio:2000} 

In this paper, we investigate structure-dynamics relations in dilute, binary mixtures of equally charged but differently sized particles by means of Brownian dynamics simulations. Since the strength of electrostatic interactions essentially depends on the number of effective charges, the interactions in such systems are practically identical between all species. Different particle sizes, however, lead to different Stokes-Einstein diffusion coefficients as short-time self-diffusion coefficients with impact both on long-time self and collective diffusion coefficients: Mediated by electrostatic interactions, the long-time dynamics of a larger species
is influenced by the presence of a smaller, more mobile species. We investigate long-time self-diffusion coefficients and, to quantify collective diffusion in mixtures of highly charged colloids, partial intermediate scattering functions where the influence of both size- and number density ratios is systematically assessed. 

For metastable, supercooled liquids, we compare different static and dynamic freezing criteria such as the Hansen-Verlet criterion\cite{hansen:1969} based on the maximum total structure factor, the dynamic Löwen criterion\cite{Loewen:1993} based on the ratio of long- to short-time self-diffusion coefficient and the Debye-Waller factor as long-time limit of the dynamic structure factor.

\section{Theoretical Background}

\subsection{Charged colloidal particles}

The interaction between charged colloidal macroions can, according to the DLVO theory, \cite{Verwey:1948} at relevant distances $r_{ij}$ be written as
\begin{align}
\dfrac{V(r_{ij})}{k_{\rm B} T} = Z_{\rm eff}^2 \lambda_{\rm B} \left( \dfrac{\re^{\kappa \sigma/2}}{1 + \kappa \sigma/2}\right)^2 \dfrac{\re^{-\kappa r_{ij}}}{r_{ij}}
\end{align}
with the thermal energy $k_{\rm B} T$, the macroions' number of effective charges $Z_{\rm eff}$, and their diameter $\sigma$.  $\lambda_{\rm B} = \re_0^2/(4\pi\epsilon_r\epsilon_0k_{\rm B} T)$ is the Bjerrum length depending on the electron charge $\re_0$ and the permittivity $\epsilon_r\epsilon_0$ of the suspending medium. 
Due to counterions present for reason of electroneutrality, the colloidal macroions interact via a screened Coulomb- or Yukawa-potential with the inverse Debye length 
\begin{align}
\kappa &= \left(4\pi\lambda_{\rm B}\sum\limits_l \,^1\rho_l z_l^2\right)^{1/2}\,,
\end{align} 
where $^1\rho_l$ is the number density of counter ions carrying $z_l$ electron charges.
Extending the single species Yukawa potential to multiple species, the pair potential reads as
\begin{align}
\dfrac{V(r_{ij})}{\kB T} &= \lambda_\mathrm{B} Z_{\mathrm{eff},i} Z_{\mathrm{eff},j} \left( \dfrac{\re^{\kappa \sigma_i/2}}{1+\kappa \sigma_i/2}\right) \left( \dfrac{\re^{\kappa \sigma_j/2}}{1+\kappa \sigma_j/2}\right) \dfrac{\re^{-\kappa r_{ij}}}{r_{ij}}.
\end{align}
Suspensions consisting of highly charged colloidal particles, due to the long-range electrostatic repulsion, self-organize to liquid-like ordered and even crystalline structures at extremely low densities \cite{Robbins:1988} with particle distances 
as large as ten times the particle diameter. For such highly dilute systems, the pairwise additive approximation of the interactions is an adequate assumption \cite{Thirumalai:1989} neglecting hydrodynamic interactions as a consequence of their extremely small volume fraction. Typical pair correlation functions $g(r)$ show, that distances comparable to the particle size do not occur: Collisions of these electrostatically interacting particles are not observed and the mutual excluded volume, unlike in hard-sphere systems, does not play a role for the self-organization. 

\subsection{Brownian dynamics simulation}

Relevant time scales for the Brownian motion of colloidal macroions are orders of magnitude larger than those of the suspending medium particles. The rapidly varying degrees of freedom of the suspending medium can be eliminated employing a projection
operator formalism \cite{zwanzig:1960,zwanzig:1961,mori:1965,mori:1965_1} leading to a generalized Fokker-Planck equation. Restricting to times larger than momentum relaxation times with typically $\tau_{p}\approx 10^{-8} {\rm s}$, such particles' equation
of motion can be described as 
\begin{align}
\vec{r}_i(t + \Delta t) =& \vec{r}_i(t) + \beta \sum\limits_{j=1}^{N} (\vec{D}_{ij} . \vec{f}_j) \Delta t \nonumber \\
&+ \sum\limits_{j=1}^{N} (\nabla_{\vec{r}_j} \vec{D}_{ij}) \Delta t + \vec{R}_i + \mathcal{O}_i(t^2)
\label{eq:displacement}
\end{align}
employing Ermak's algorithm \cite{Ermak:1975, Ermak:1975_1, Ermak:1978}.
Here, $\vec{r}_i(t)$ is the position of particle $i$ at time $t$, $\beta=(\kBT)^{-1}$ the inverse thermal energy, and $\m{D}_{ij}$ the diffusion tensor. The force acting on particle $j$ is denoted as $\vec{f}_j$ and $\vec{R}_i$ is a random displacement caused by the rapidly fluctuating interactions between the suspending medium and the colloidal macroions with the properties
\begin{align}
\left\langle \vec{R}_i(\tau) \right\rangle &= 0
\end{align}
and
\begin{align}
\left\langle \vec{R}_i(\tau) . \vec{R}_i(\tau+\Delta t) \right\rangle = 6 D_0 \Delta t\,
\end{align}
neglecting hydrodynamic interactions. In pairwise additive approximation, the force $\vec{f}_j$ acting on particle $j$ reads as
\begin{align}
\vec{f}_j = - \sum \nabla_{\vec{r}_i} V(\vec{r}_{ij})
\end{align}
with $V(\vec{r}_{ij})$ being the potential of two macroions with distance vector $\vec{r}_{ij}$.
Neglecting hydrodynamic interactions in highly dilute suspensions, the diffusion tensor $\mathbf{D}_{ij}$ is approximated as $\mathbf{D}_{ij}=D_{0,i} \mathbf{I}$ with $\mathbf{I}$ denoting matrix identity and
\begin{align}
D_{0,i} &= \dfrac{k_{\rm B} T}{3\pi\eta \sigma_i}
\end{align}
the Stokes-Einstein diffusion coefficient of particle $i$ depending on the viscosity $\eta$ of the suspending medium and the particle diameter $\sigma_i$. In absence of hydrodynamic interactions, the derivative $\nabla_{\vec{r}_j}\mathbf{D}_{ij}=0$ vanishes.

\subsection{Dynamic properties}
The time-dependent self-diffusion coefficient $D_{{\rm S},i}(t)$ of a particle $i$ is accessible via its mean-square displacement as
\begin{align}
D_{{\rm S},i}(t) &= \dfrac{1}{6}\dfrac{\partial}{\partial t} \left\langle\left[ \mathbf{r}_i(\tau)-\mathbf{r}_i(\tau+t)\right]^2 \right\rangle_\tau\,.
\end{align}
Both, the short-time limit $D_0$ and the long-time limit $D_{\rm S}^{\rm (L)}$ are time-independent. The short-time self-diffusion coefficient is the Stokes-Einstein diffusion coefficient. The reduced self-diffusion coefficient 
\begin{align}
D_{\mathrm{S},i}^{\mathrm{red}}(t) &= \dfrac{D_{\mathrm{S},i}(t)}{D_{0,i}},
\end{align}
as a size-independent quantity allows to analyze  the time-dependence of the self-diffusion in mixtures.

In an alternative way, the time-dependent self-diffusion of particle $i$ is accessible from the van Hove function \cite{Hopkins:2010}
\begin{align}
G(\vec{r}_{ij},t) &= \dfrac{1}{N} \Braket{\sum\limits_{i=1}^{N}\sum\limits_{j=1}^{N} \delta\left[\vec{r} - \vec{r}_i(t) + \vec{r}_j(0)\right]}\\
&= G_\mathrm{S}(\vec{r}_{ii}, t) + G_\mathrm{D}(\vec{r}_{ij}, t)
\end{align}
with $G_\mathrm{S}$ denoting its self and $G_\mathrm{D}$ its collective part. 

In a mixture of $N_{\rm C}$ species for each species $\alpha$ a self correlation function $G_{{\rm S},\alpha}(\vec{r}_{ii},t)$ exists which
obeys the differential equation
\begin{align}
\label{eq:van_Hove_Self_Diff}
\dfrac{\partial G_{{\rm S},\alpha}(\vec{r}_{ii}, t)}{\partial t} &= D_{{\rm S},\alpha}(t)\nabla^2 G_{{\rm S},\alpha}(\vec{r}_{ii}, t)
\end{align}
with the solution
\begin{align}
G_{{\rm S},\alpha}(r_{ii}, t) &= (4\pi D_{{\rm S},\alpha}(t) t)^{-3/2}\exp\left({-\frac{r_{ii}^2}{4D_{{\rm S},\alpha}(t) t}}\right)
\label{eq:self_van_hove_approx}
\end{align}
where in the limit $t\to \infty$ the long-time self-diffusion coefficient $D_{{\rm S},\alpha}(t\to\infty) = D_{{\rm S},\alpha}^{\rm (L)}$ results.

In such a mixture, $N_{\rm C}\left(N_{\rm C}+1\right)/2$ different, distinct van Hove functions $G_{{\rm D},\alpha\beta}(\mathbf{r}_{j\beta}-\mathbf{r}_{i\alpha},t)$
\begin{align}
G_{{\rm D},\alpha\beta}(\vec{r},t) &= \dfrac{1}{\left(N_\alpha N_\beta\right)^{1/2}} \Braket{\sum\limits_{i}\sum\limits_{j\neq i} \delta\left[\vec{r}- \vec{r}_{j\beta}(\tau+t) + \vec{r}_{i\alpha}(\tau)\right]}_\tau.
\end{align}
exist and are accessible from the trajectories.

Analogously to Eq. \eqref{eq:van_Hove_Self_Diff}, the distinct part of the van Hove function obeys the differential equation
\begin{align}
\label{eq:van_Hove_Dist_Diff}
\dfrac{\partial G_{{\rm D},\alpha\beta}(\vec{r}_{ij}, t)}{\partial t} &= D_{{\rm C},\alpha\beta}(t)\nabla^2 G_{{\rm D},\alpha\beta}(\vec{r}_{ij}, t)
\end{align}
which reads as
\begin{align}
\label{eq:van_Hove_Dist_reciprokal_space}
\dfrac{\partial G_{{\rm D},\alpha\beta}(\mathbf{Q}, t)}{\partial t} &= -D_{{\rm C},\alpha\beta}(t)Q^2 G_{{\rm D},\alpha\beta}(\mathbf{Q}, t)
\end{align}
in reciprocal space. The van Hove function $G_{{\rm D},\alpha\beta}(\mathbf{Q}, t)$ is related to the intermediate scattering function $S_{\alpha\beta}(\mathbf{Q},t)$ as
\begin{align}
S_{\alpha\beta}(\mathbf{Q}, t)=G_{{\rm D},{\alpha\beta}}(\mathbf{Q},t)+\delta_{\alpha\beta}
\end{align}
with $\delta_{\alpha\beta}$ denoting the Kronecker symbol.

In a binary mixture, three partial intermediate scattering functions are accessible from the trajectories. Normalizing by the static structure factors $S_{\alpha\beta}(\mathbf{Q})=S_{\alpha\beta}(\mathbf{Q},t=0)$, reduced intermediate scattering functions
\begin{align}
S_{{\rm D},\alpha\beta}^{\mathrm{red}}(\mathbf{Q},t) = \dfrac{S_{\alpha\beta}(\mathbf{Q},t) - \delta_{\alpha\beta}}{S_{\alpha\beta}(\mathbf{Q},0) - \delta_{\alpha\beta}}
\end{align}
are obtained. Due to spherical symmetry, for liquid like structures, these quantities depend only on the modulus $Q=\Vert \mathbf{Q}\Vert$ of the wave vector.

To identify a freezing transition in interacting systems, different criteria are described in the literature. The Hansen-Verlet criterion \cite{hansen:1969} is based on the maximum of the static structure factor and identifies a frozen state for systems with $\max[S(Q)] \gtrsim 2.85$. Löwen \etal\, suggest a dynamical freezing criterion based on the reduced long-time self-diffusion coefficient, to identify a frozen state when $D_{\rm S}^{\rm (L)}/D_0 \lesssim 0.098$ \cite{Loewen:1993, Torres:2023}. A frequently used criterion to characterize the glass transition is a non-zero Debye-Waller factor $S(Q,t\to\infty)/S(Q,0)$ indicating a frustrated
$\alpha$-process.\cite{goetze:1992, Bermejo:1993, goetze:2009, Amann:2012, Ghosh:2023}

\subsection{Computational details}

We performed Brownian dynamics simulations employing the previously described and well known Ermak-McCammon algorithm \cite{Ermak:1978}, neglecting hydrodynamic interactions in a first approach.
The mean-square displacement $\braket{\Delta r_i^2(t)}$ and partial distinct van Hove functions are calculated from the simulation trajectories. By Fourier transform, from the latter quantities intermediate scattering functions $S(Q,t)$ are obtained. The parameters used for Brownian dynamics simulations are compiled in Tables \ref{tab:const_params} and \ref{tab:consts_example}.

\begin{table}[h]
\renewcommand{\arraystretch}{1.2}
\centering
\caption{Simulation parameters and properties of the suspending medium (water) used in Brownian dynamics simulations of binary colloidal Yukawa systems.}
\label{tab:const_params}
\begin{tabular}{ll}
\hline\hline
Parameter & SI units \\
\hline
Temperature & $T = \SI{298.15}{\kelvin}$\\
Box length & $L_{\rm B} = \SI{2.52e-5}{\meter}$\\
Total number density\phantom{X} & $^1\rho_{\rm tot} = \SI{1e18}{\per\cubic\meter}$\\
Time base & $\tau = \SI{1e-3}{\second}$\\
Time step & $\Delta t = \SI{2.0e-6}{\second}$\\
Viscosity & $\eta = \SI{8.9e-4}{\pascal\second}$\\
Relative permittivity & $\epsilon_{r} = \num{78.3}$\\
Bjerrum length & $\lambda_{\rm B} = \SI{7.160e-10}{\meter}$\\
Particle number & $N_{\rm P} = 16000$\\
\hline\hline
\end{tabular}
\end{table}

The diameter $\sigma_{\rm A} = \SI{100}{\nano\meter}$ is kept constant during all simulations, while $\sigma_{\rm B}$ is varied from \SI{100}{\nano\meter} to \SI{020}{\nano\meter} to obtain
binary mixtures including a virtually binary mixture with $\sigma_{\rm A}=\sigma_{\rm B}=\SI{100}{\nano\meter}$.
The number of effective charges of both species is varied in the range of $Z_{\rm eff, A} = Z_{\rm eff, B} = \numrange{300}{600}$. To investigate the influence of relative particle number densities $^1\rho_{\rm B}/\,^1\rho_{\rm A}$, three different ratios $^1\rho_{\rm B}/\,^1\rho_{\rm A}=\{1, 2, 4\}$ are simulated.
Additionally, in Table \ref{tab:consts_example} the reduced interaction parameters
\begin{align}
A_{ij} &= \lambda_\mathrm{B} Z_{\mathrm{eff},i} Z_{\mathrm{eff},j} \left( \dfrac{\re^{\kappa \sigma_i/2}}{1+\kappa \sigma_i/2}\right) \left( \dfrac{\re^{\kappa \sigma_j/2}}{1+\kappa \sigma_j/2}\right)
\end{align}
are exemplarily compiled for $\sigma_{\rm A}=\SI{100}{\nano\meter}$ and $\sigma_{\rm B}=\SI{50}{\nano\meter}$ with $Z_{\rm eff,A}=Z_{\rm eff,B}=400$. 
The interaction parameters differ by less than one percent at identical number of effective charges for all simulated particle sizes.
Depending on the particle sizes, the volume fractions of the investigated systems are in the range of $4.2\times 10^{-6}\le \varphi \le 5.2\times 10^{-4}$.

\begin{table}
\renewcommand{\arraystretch}{1.2}
\centering
\caption{Particle-dependend parameters used for Brownian dynamics simulations of binary mixtures, exemplarily compiled for particles with diameter $\sigma_{\rm A}= \SI{100}{\nano\meter}$ and $\sigma_{\rm B} = \SI{50}{\nano\meter}$ with identical number of effective charges $Z_{\rm eff, A} = Z_{\rm eff, B} = 400$ and identical number densities $^1\rho_{\rm A} = \,^1\rho_{\rm B} = \SI{5e17}{\per\cubic\meter}$.}
\label{tab:consts_example}
\resizebox{\columnwidth}{!}{
\begin{tabular}{ll}
\hline\hline
SI units & Reduced units\\
\hline
$L_{\rm B} = \SI{2.52e-5}{\meter}$ & $L^* = L_{\rm B} /\sigma_{\rm A} = 252.0$\\
$^1\rho_{\rm tot} = \SI{1e18}{\per\cubic\meter}$ & $^1\rho^*_{\rm tot} = \,^1\rho_{\rm tot} \sigma_{\rm A}^3 = 0.001$\\
$\Delta t = \SI{2e-6}{\second}$ & $\Delta t^* = \Delta t / \tau = 0.002$\\
$\kappa = \SI{1.897e5}{\per\meter}$ & $\kappa^* = \kappa \sigma_{\rm A} = 0.1897$\\
$D_{\rm 0, A} = \SI{4.908e-12}{\square\meter\per\second}$ & $D_{\rm 0, A}^* = D_{\rm 0, A} \tau / \sigma_{\rm A}^2 = 0.4908$\\
$D_{\rm 0, B} = \SI{9.815e-12}{\square\meter\per\second}$ & $D_{\rm 0, B}^* = D_{\rm 0, B} \tau / \sigma_{\rm A}^2 = 0.9815$\\
$A_{\rm AA} = \SI{1.155e-4}{\meter}$ & $A_{\rm AA}^* = A_{\rm AA} / \sigma_{A} = 1155$\\
$A_{\rm AB} = \SI{1.151e-4}{\meter}$ & $A_{\rm AB}^* = A_{\rm AB} / \sigma_{A} = 1151$\\
$A_{\rm BA} = A_{\rm AB}$ & $A_{\rm BA}^* = A_{\rm AB}^*$\\
$A_{\rm BB} = \SI{1.148e-4}{\meter}$& $A_{\rm BB}^* = A_{\rm BB} / \sigma_{A} = 1148$\\
\hline\hline
\end{tabular}
}
\end{table}

Starting from a bcc structure, in ten successive simulations, each with \num{e4} time steps, with drastically reduced number of effective charges of $Z_{\rm eff, A} = Z_{\rm eff, B} = 50$, a gaseous configuration is prepared. Using this disordered configuration, in ten successive simulation runs of each \num{e4} time steps liquid-like ordered systems consisting of particles with the respective number of effective charges are prepared. The convergence of equilibration is monitored by comparison of static pair correlation functions computed in each run. After equilibration, a production run of \num{2e6} time steps is performed.

From its trajectory, the mean-square displacement and the distinct van Hove functions are computed.
To estimate uncertainties of the results, twelve independent production runs are evaluated for each system.
The uncertainties are given as triple standard deviations of the results.

\section{Results and Discussion}

\begin{figure}[h!]
\centering
\includegraphics[width=1.0\columnwidth]{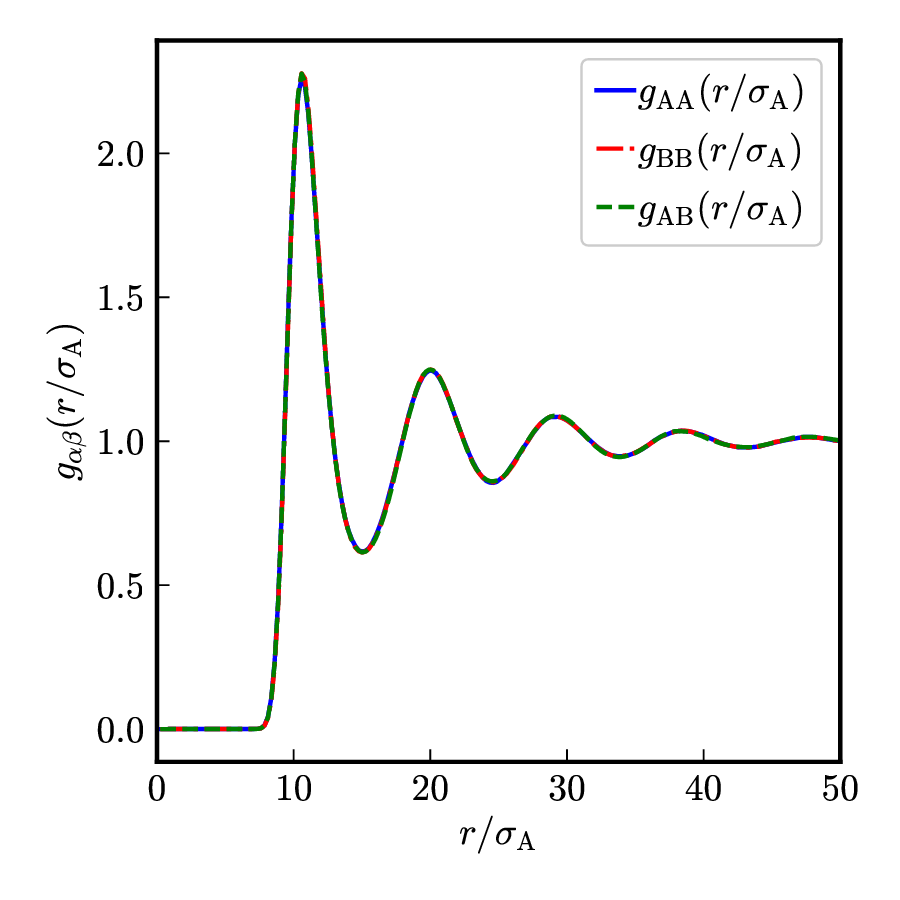}
\caption{Partial pair correlation functions $g_{\alpha\beta}(r /\sigma_\mathrm{A})$ of a binary suspension of Yukawa particles. The suspension consists of a large species with the diameter $\sigma_\mathrm{A} = \SI{100}{\nano\meter}$ and a smaller species with the diameter $\sigma_\mathrm{B} = \SI{50}{\nano\meter}$. The number of effective charges of both species is $Z_\mathrm{eff, A} = Z_{\rm eff, B} = 400$ at identical number densities $^1\rho_\mathrm{A}=\,^1\rho_\mathrm{B} = \SI{5e17}{\per\cubic\meter}$.}
\label{fig:static_pair_correlation}
\end{figure}

In the present work, we investigate binary colloidal suspensions of equally charged, but differently sized particles, interacting via a repulsive, screened electrostatic or Yukawa-potential.
The interaction is primarily governed by the number of effective charges of the particles, the permittivity of the suspending medium, the temperature, and the mean interparticle distance depending on the
total number density. Therefore, the interaction only slightly depends on the particles' size and is neglectable.
Due to practically identical interactions between all species in time average, as visible in Fig. \ref{fig:static_pair_correlation}, the partial pair correlation functions  $g_{\rm AA}(r)$, $g_{\rm AB}(r)$, and $g_{\rm BB}(r)$ are identical.

\subsection{Self dynamics}

\begin{figure*}[t]
\centering
\includegraphics[width=0.95\textwidth]{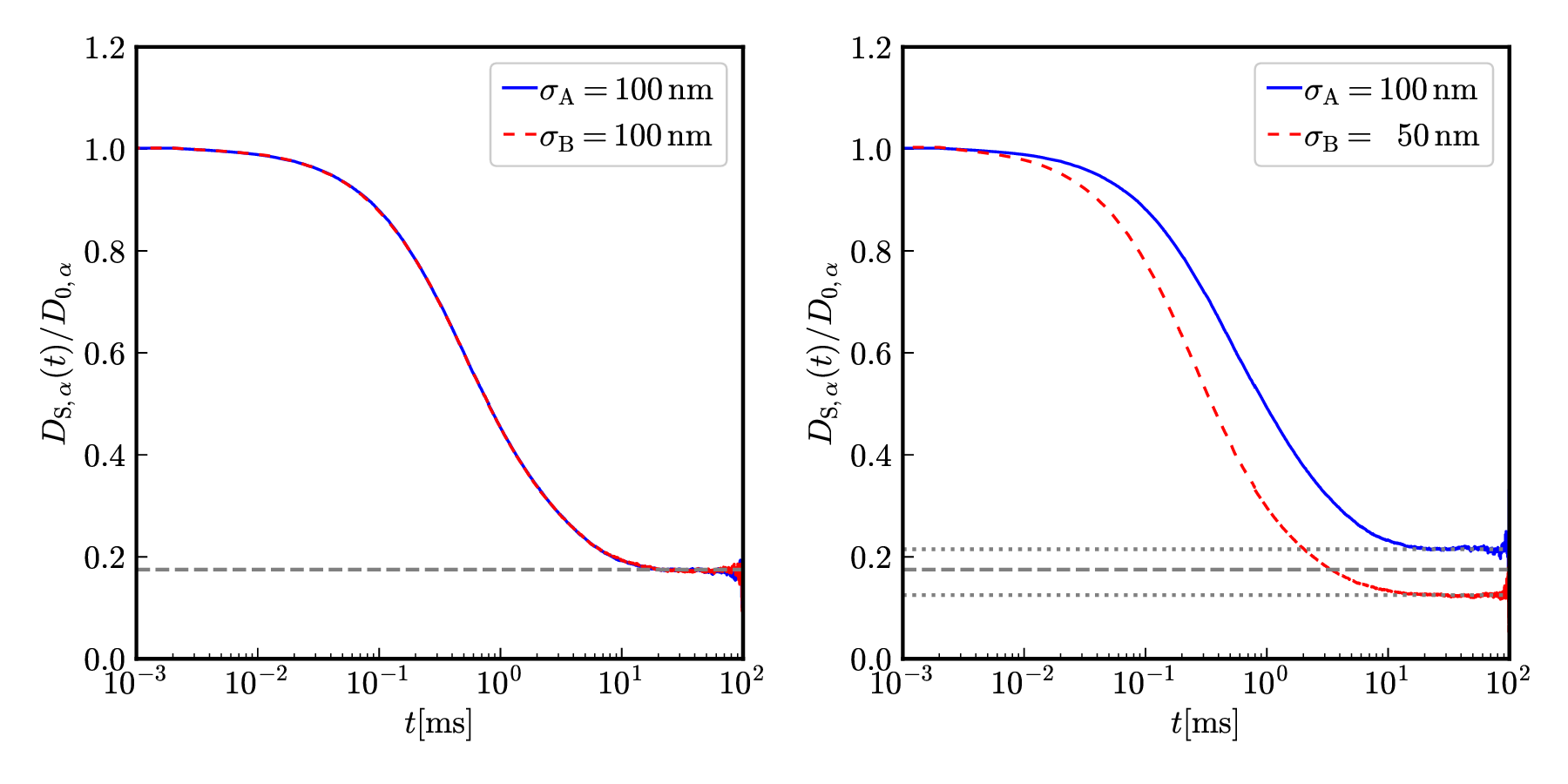}
\caption{Reduced time-dependent self-diffusion coefficients $D_{{\rm S},\alpha}(t) / D_{0,\alpha}$ of a virtually binary suspension of identically sized species with diameters $\sigma_{\rm A} = \sigma_{\rm B} = \SI{100}{\nano\meter}$, identical number of effective charges $Z_\mathrm{eff, A} = Z_{\rm eff, B} = 400$ and number densities $^1\rho_\mathrm{A}=\,^1\rho_\mathrm{B} = \SI{5e17}{\per\cubic\meter}$ (left). Corresponding reduced time-dependent self-diffusion coefficients of a binary colloidal mixture consisting of differently sized species with $\sigma_{\rm A} = \SI{100}{\nano\meter}$ and $\sigma_{\rm B} = \SI{50}{\nano\meter}$, but still identical number of effective charges $Z_{\rm eff, A} = Z_{\rm eff, B} = 400$  and number densities $^1\rho_\mathrm{A}=\,^1\rho_\mathrm{B} = \SI{5e17}{\per\cubic\meter}$ (right). The corresponding long-time limits are indicated as a dashed line for the virtually binary mixture and as dotted lines for the binary mixture.}
\label{fig:dsl_time_dependence}
\end{figure*}

In Fig. \ref{fig:dsl_time_dependence}, the reduced time-dependent self-diffusion coefficients in a virtually binary mixture of identical particles (lhs) and differently sized, but equally charged particles (rhs)
are displayed. In the virtually binary mixture of identical particles, the time-dependence of the self-diffusion coefficients is identical. Opposite, in presence of a differently sized, smaller species, the reduced long-time self-diffusion coefficient $D_{\rm S,A}^{\rm (L)}/D_{0,{\rm A}}$ is enhanced while that $D_{\rm S,B}^{\rm (L)}/D_{0,{\rm B}}$ is reduced. In the short-time limit, consistently, the respective Stokes-Einstein diffusion coefficients are approached. For intermediate times, a subdiffusive motion leading to formal time-dependent self-diffusion coefficients is observed.
In Table \ref{tab:dsl_pure}, the reduced long-time self-diffusion coefficients of one-component suspensions are compiled in dependence on the particle diameter $\sigma$ and the number of effective charges $Z_{\rm eff}$.

\begin{figure}[h!]
\centering
\includegraphics[width=1.0\columnwidth]{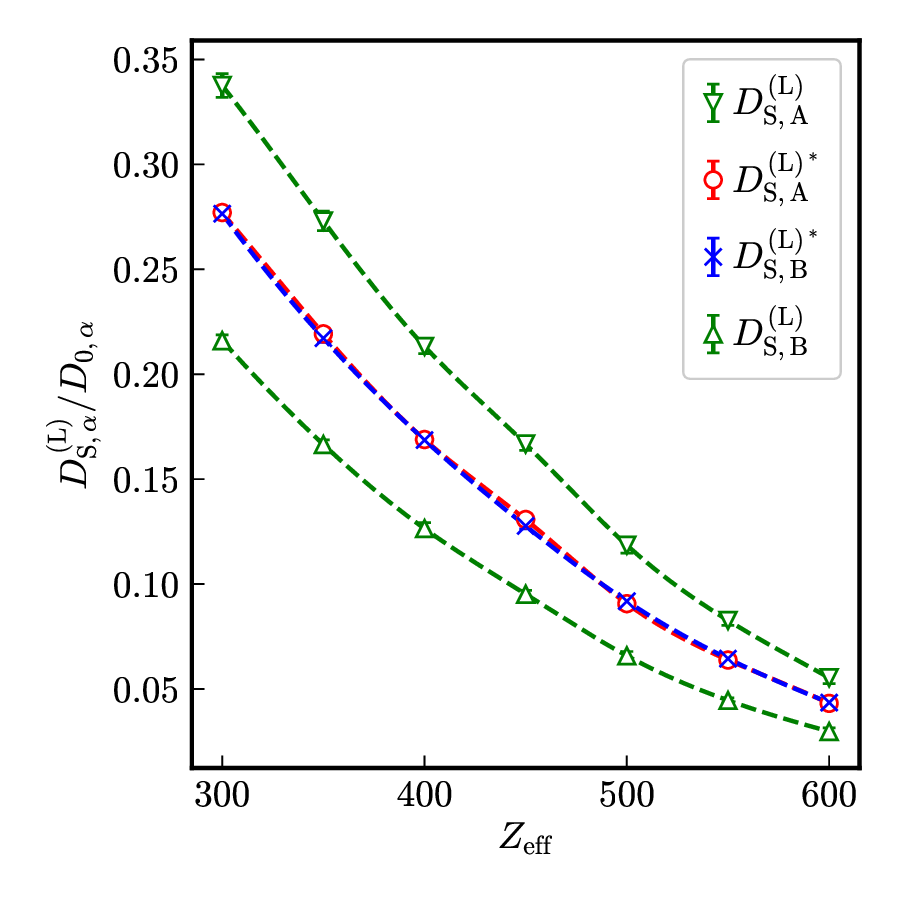}
\caption{Reduced long-time self-diffusion coefficients of a binary mixture in dependence on the number of effective charges $Z_\mathrm{eff}$. Those of the large species are depicted as open upper triangles while those of the small species are represented by open lower triangles. For comparison, the values for one-component systems of both species with the same total number density are displayed as open circles and crosses. The dashed lines are cubic splines as a guide to the eye.}
\label{fig:dsl_binary_vs_zeff}
\end{figure}

In contrast to the effective one-component system, different Stokes-Einstein diffusion coefficients of a mixture's constituents lead to a different time-dependence of their self-diffusion coefficients: In the short-time limit,
both self-diffusion coefficients approach the respective Stokes-Einstein diffusion coefficient, while in the long-time limit different reduced long-time self-diffusion coefficients $D_{{\rm S},\alpha}^{\rm (L), red}$ are observed.
The influence of the identical number of effective charges $Z_{\rm eff,A}=Z_{\rm eff,B}$ is exemplarily visible in Fig. \ref{fig:dsl_binary_vs_zeff} for a binary suspension consisting of particles with diameter $\sigma_{\rm A}= \SI{100}{\nano\meter}$ and $\sigma_{\rm B} = \SI{50}{\nano\meter}$ at identical particle number densities $^1\rho_{\rm A} = \,^1\rho_{\rm B} = \SI{5e17}{\per\cubic\meter}$. For comparison,
the self-diffusion coefficients in a virtually binary mixture of identical particles, i.e., an effective one-component system, is depicted as well. 
With increasing number of effective charges and therewith increasing interactions, the long-time self-diffusion coefficient non-linearly decreases.

Generally, the long-time self-diffusion coefficients of a larger species with smaller Stokes-Einstein diffusion coefficient are enhanced by the presence of a smaller species with correspondingly larger Stokes-Einstein
diffusion coefficients and vice versa. The differences of long-time self-diffusion coefficients decrease at identical ratios of sizes and number densities with increasing strength of electrostatic interaction.

\begin{figure}[h!]
\centering
\includegraphics[width=1.0\columnwidth]{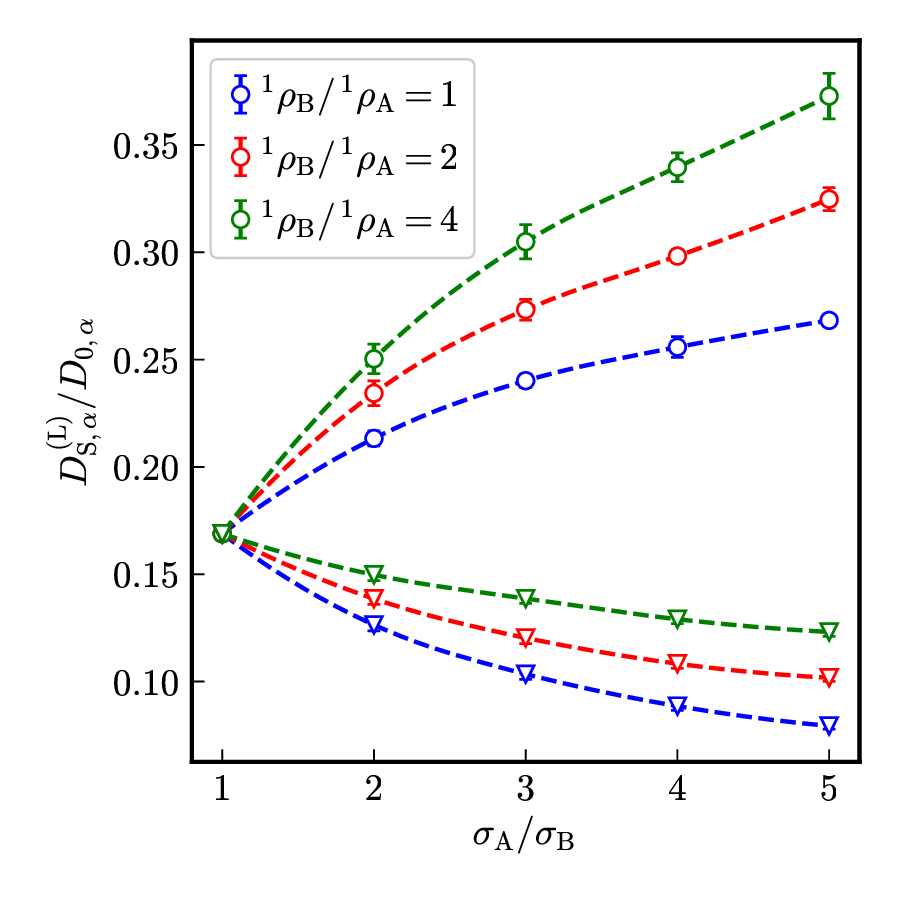}
\caption{Reduced long-time self-diffusion coefficients $D_{\mathrm{S},\alpha}^{(\mathrm{L})}/D_{0,\alpha}$ for a binary suspension with number of effective charges $Z_\mathrm{eff} = 400$ in dependence on the diameter ratio $\sigma_\mathrm{A}/\sigma_\mathrm{B}$ and the ratio of their number densities $^1\rho_\mathrm{B}/\,^1\rho_\mathrm{A}$. The dashed lines are cubic splines as a guide to the eye.}
\label{fig:dsl_sig_rho}
\end{figure}

The reduced long-time self-diffusion coefficients of binary suspensions in dependence on the number of effective charges $Z_{\rm eff}$, the particle-size ratio $\sigma_{\rm A} / \sigma_{\rm B}$ and the ratio of number densities $^1\rho_{\rm B} / \,^1\rho_{\rm A}$ are compiled in Table \ref{tab:dsl_data}.
As visible in Fig. \ref{fig:dsl_sig_rho}, with increasing particle-size ratio, the acceleration and deceleration of reduced long-time self-diffusion coefficients increases, too.
In suspensions with different number densities $^1\rho_{\rm A} \neq \,^1\rho_{\rm B}$, reduced long-time self-diffusion coefficients of the minority component change more than those of the majority component.

\begin{figure}[h!]
\centering
\includegraphics[width=1.0\columnwidth]{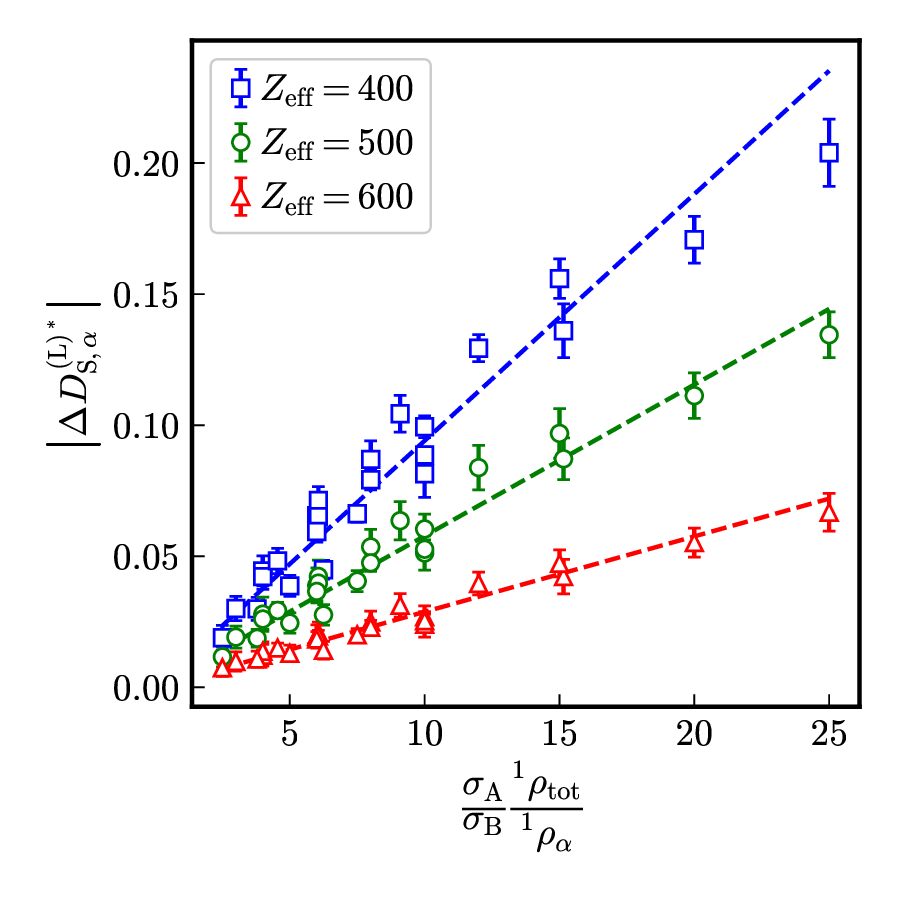}
\caption{Modulus of the reduced excess long-time self-diffusion coefficient in dependence on the ratio $\left(\sigma_{\rm A}\,^1\rho_{\rm tot}\right) / \left(\sigma_{\rm B} \,^1\rho_{\alpha}\right)$. The dashed lines are linear fits to the data.
}
\label{fig:acceleration_normed}
\end{figure}

\begin{figure}[h!]
\centering
\includegraphics[width=1.0\columnwidth]{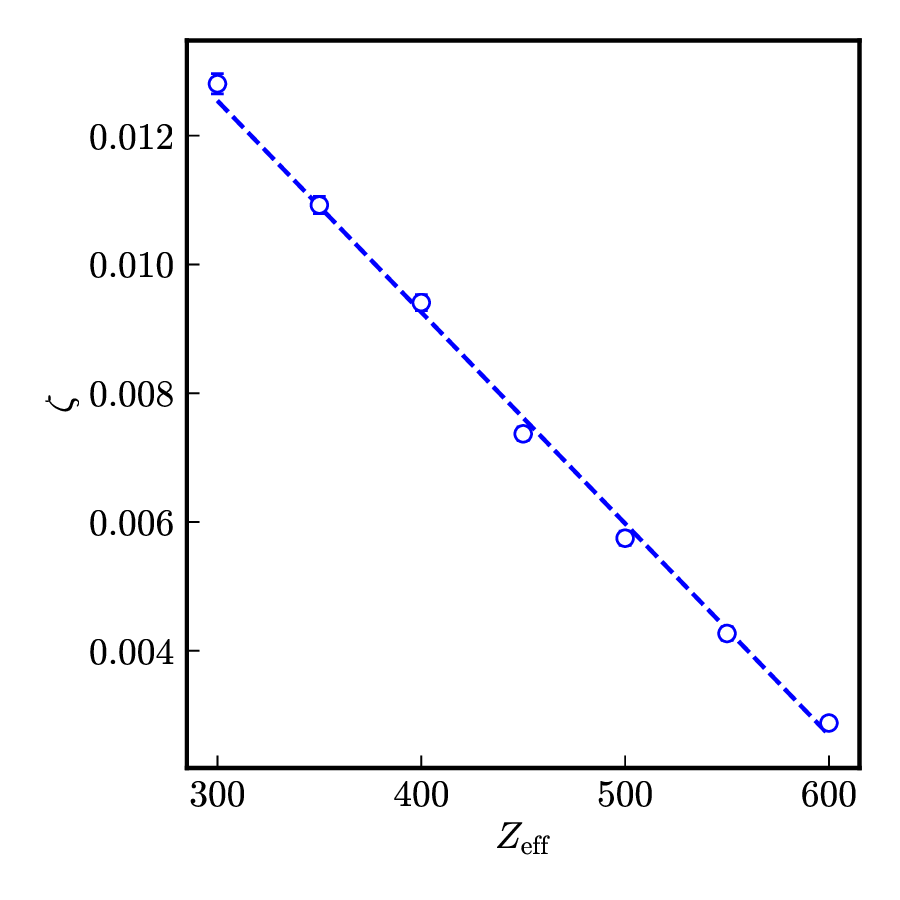}
\caption{Dependency of the coupling constant $\zeta$ on the number of effective charges $Z_{\rm eff}$. The dashed line is a linear fit to the data.\vspace{1cm}}
\label{fig:coupling_constant}
\end{figure}

Let us define a reduced excess long-time self-diffusion coefficient
\begin{align}
\Delta D_{{\rm S},\alpha}^{\rm (L)^*} = \dfrac{D_{{\rm S},\alpha}^{\rm (L)} - D_{{\rm S},\alpha}^{\rm (L)^*}}{D_{0,\alpha}} 
\end{align}
as difference of the reduced long-time self-diffusion coefficient of species $\alpha$ in a mixture with total number density $^1\rho_{\rm tot}$ to that of species $\alpha$ in a one-component system with $^1\rho_\alpha=\,^1\rho_{\rm tot}$. As visible in Fig. \ref{fig:acceleration_normed}, for sufficiently large electrostatic interaction, a nearly linear dependence of the modulus of this reduced excess long-time self-diffusion coefficients on the ratio $\left(\sigma_{\rm A}\,^1\rho_{\rm tot}\right) / \left(\sigma_{\rm B} \,^1\rho_{\alpha}\right)$ is observed. 
The slope of this nearly linear dependence can be considered as a dynamic coupling constant $\zeta$. As visible from Fig. \ref{fig:coupling_constant}, the dynamic coupling constant linearly depends on the number of effective charges (Table \ref{tab:coupling_constant}). 

\begin{table}[h!]
\renewcommand{\arraystretch}{1.2}
\centering
\caption{Coupling constant $\zeta$ in dependence on the number of effective charges $Z_\mathrm{eff}$.}
\label{tab:coupling_constant}
\begin{tabular}{lccccccc}
\hline
\hline
$Z_\mathrm{eff}$ & 300 & 350 & 400 & 450 & 500 & 550 & 600\\
\hline
$\zeta \times 10^{-2}$ & $1.28_2$ & $1.092_{14}$ & $0.941_{13}$ & $0.737_{11}$ & $0.555_{9}$ & $0.391_{8}$ & $0.249_{7}$\\
\hline
\hline
\end{tabular}
\end{table}

\subsection{Collective dynamics}

In addition to the self-dynamics, we investigate the collective dynamics employing intermediate scattering functions obtained by Fourier transform of distinct van Hove functions. In the here investigated liquid-like 
systems, all partial time-dependent structure factors decay to zero in the long-time limit. We quantify the collective short-time behavior by stretched exponentials $\exp(-at)^b$ as a heuristic approach exemplarily at
the wave-vector $Q_{\rm max}$ at the maximum of the static structure factor.

\begin{figure}[h]
\centering
\includegraphics[width=1.0\columnwidth]{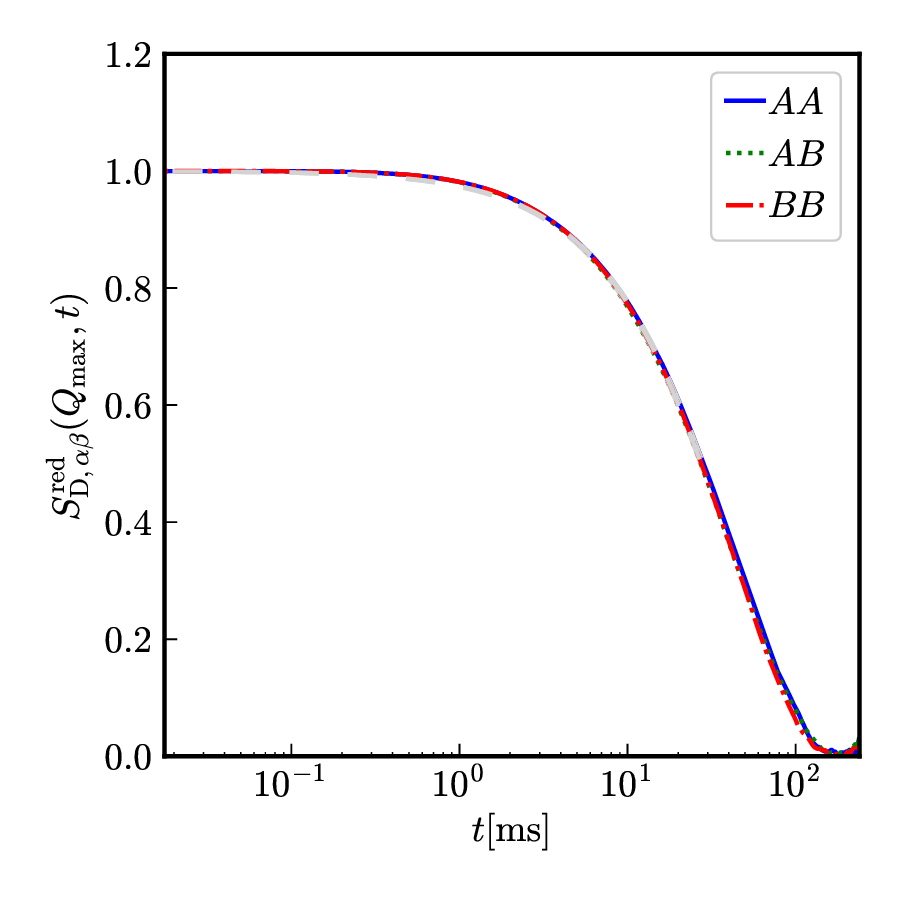}
\caption{Reduced time-dependent, partial distinct structure factors $S_{{\rm D},\alpha\beta}^{\rm red}(Q_{\rm max}, t)$ of a virtually binary suspension of identical particles with diameter $\sigma = \SI{100}{\nano\meter}$ at identical number density $^1\rho_{\rm A}=\,^1\rho_{\rm B}=\SI{5e17}{\per\cubic\meter}$ with $Z_{\rm eff} = 400$ effective charges. The stretched exponentials are displayed as dashed grey lines.}
\label{fig:reduced_sqtd_pure}
\end{figure}

Let us first consider a virtually binary suspension consisting of identical particles. In Fig. \ref{fig:reduced_sqtd_pure}, the reduced partial, distinct intermediate scattering functions $S_{{\rm D},\alpha\beta}^{\rm red}(Q_{\rm max}, t)$ are shown for a suspension of colloidal macroions with diameter $\sigma_{\rm A} = \sigma_{\rm B} = \SI{100}{\nano\meter}$,  $Z_{\rm eff, A} = Z_{\rm eff, B} = 400$ effective charges at identical number densities $^1\rho_{\rm A} = \,^1\rho_{\rm B} = \SI{5e17}{\per\cubic\meter}$. Here, the functional form of decay is identical for all partial intermediate scattering functions.

The relaxation rates $a$ and exponents $b$ of stretched exponentials observed for virtually binary mixtures are compiled in Table \ref{tab:alpha_beta_pure}. These parameters are extracted from fits covering correlation times until $S_{\alpha\beta}(Q,t)=S_{\alpha\beta}(Q,0)/2$ is reached.

\begin{figure}
\centering
\includegraphics[width=1.0\columnwidth]{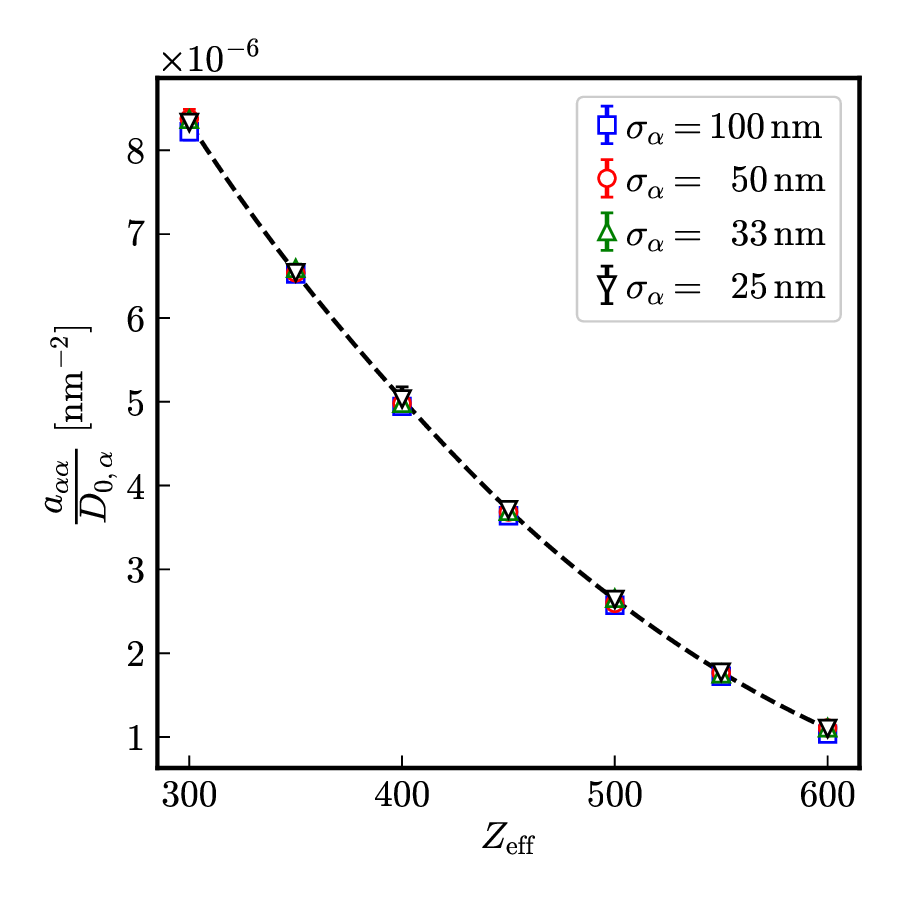}
\caption{Relaxation rates $a_{\alpha\alpha}$ normalized to the respective Stokes-Einstein diffusion coefficient $D_{0,\alpha}$  at $Q_{\rm max}$ in virtually binary mixtures of identical particles in dependence on their number of effective charges $Z_{\rm eff}$ for different particle sizes $\sigma_\alpha$.  The dashed line is a cubic spline as a guide to the eye. 
\label{fig:alpha_vs_zeff_pure}
}
\end{figure}

In Fig. \ref{fig:alpha_vs_zeff_pure}, the relaxation rates $a$ in effective one-component systems are displayed in dependence on the number of effective charges $Z_{\rm eff}$ for different diameters. Normalized to the
respective Stokes-Einstein diffusion coefficients $D_{0,\alpha}$ a universal reduced relaxation rate, decreasing with the number of effective charges $Z_{\rm eff}$ is obtained.
The exponent $b$ decreases in first approximation linearly with increasing number of effective charges and is independent of the particle size.

\begin{figure}
\centering
\includegraphics[width=1.0\columnwidth]{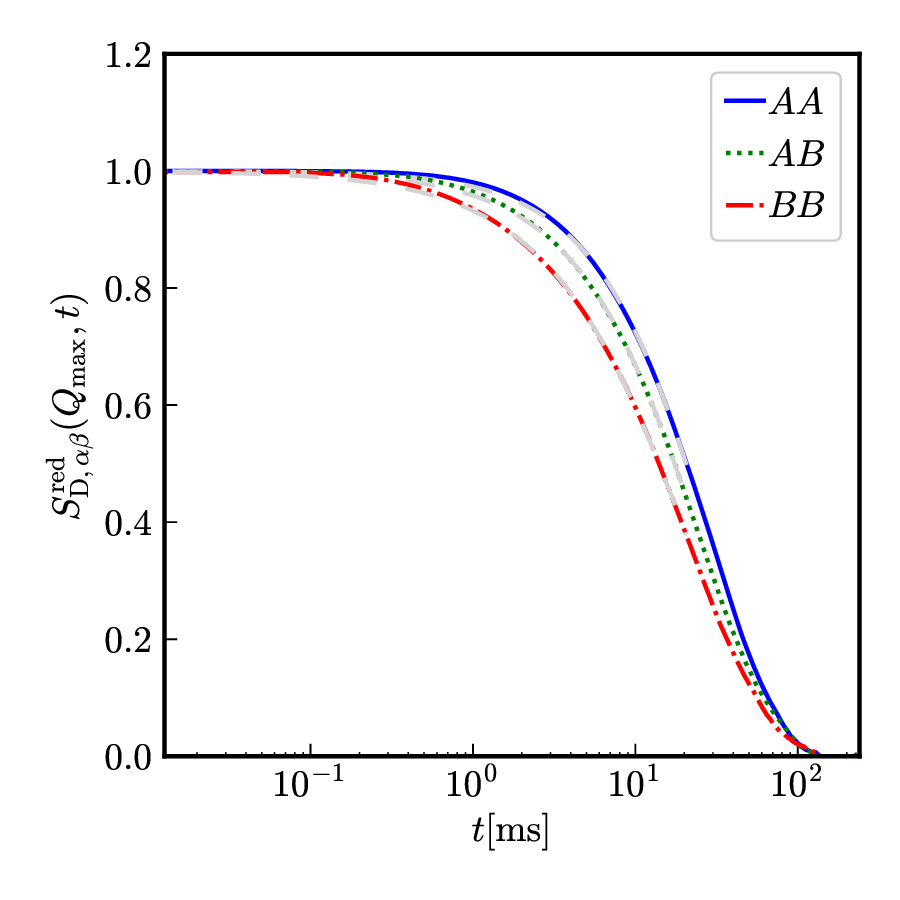}
\caption{Partial intermediate scattering functions $S_{{\rm D},\alpha\beta}^{\rm red}(Q_{\rm max}, t)$ of a binary colloidal suspension of particles with diameters $\sigma_{\rm A} = \SI{100}{\nano\meter}$ and  $\sigma_{\rm B} = \SI{33}{\nano\meter}$  with identical number of effective charges $Z_{\rm eff, A} = Z_{\rm eff, B} = 400$ at identical number densities $^1\rho_{\rm A} = \,^1\rho_{\rm B} = \SI{5e17}{\per\cubic\meter}$. The dashed grey lines are stretched exponential fits of the data.}
\label{fig:reduced_sqtd_binary}
\end{figure}

In a binary mixture of differently sized but still equally charged particles, the time-dependence of partial intermediate scattering functions differs as exemplarily shown in Fig. \ref{fig:reduced_sqtd_binary} for
particles with diameters $\sigma_{\rm A}=100\,{\rm nm}$ and $\sigma_{\rm B}=33\,{\rm nm}$ with $Z_{\rm eff,A}=Z_{\rm eff,B}=400$ effective charges at identical number densities $^1\rho_{\rm A}=\,^1\rho_{\rm B}=\SI{5e17}{\per\cubic\meter}$. As expected, the decay of $S_{\rm AA}(Q,t)$ is the slowest and that of $S_{\rm BB}(Q,t)$ the fastest. The relaxation of the
correlation between both species $S_{\rm AB}(Q,t)$ is in between $S_{\rm AA}(Q,t)$ and $S_{\rm BB}(Q,t)$. The relaxation rates $a_{\alpha\beta}$ and stretching exponents $b_{\alpha\beta}$ for all investigated binary mixtures are compiled in Tables \ref{tab:alpha_beta_mix_1}, \ref{tab:alpha_beta_mix_2} and \ref{tab:alpha_beta_mix_4}.

\begin{figure}
\centering
\includegraphics[width=1.0\columnwidth]{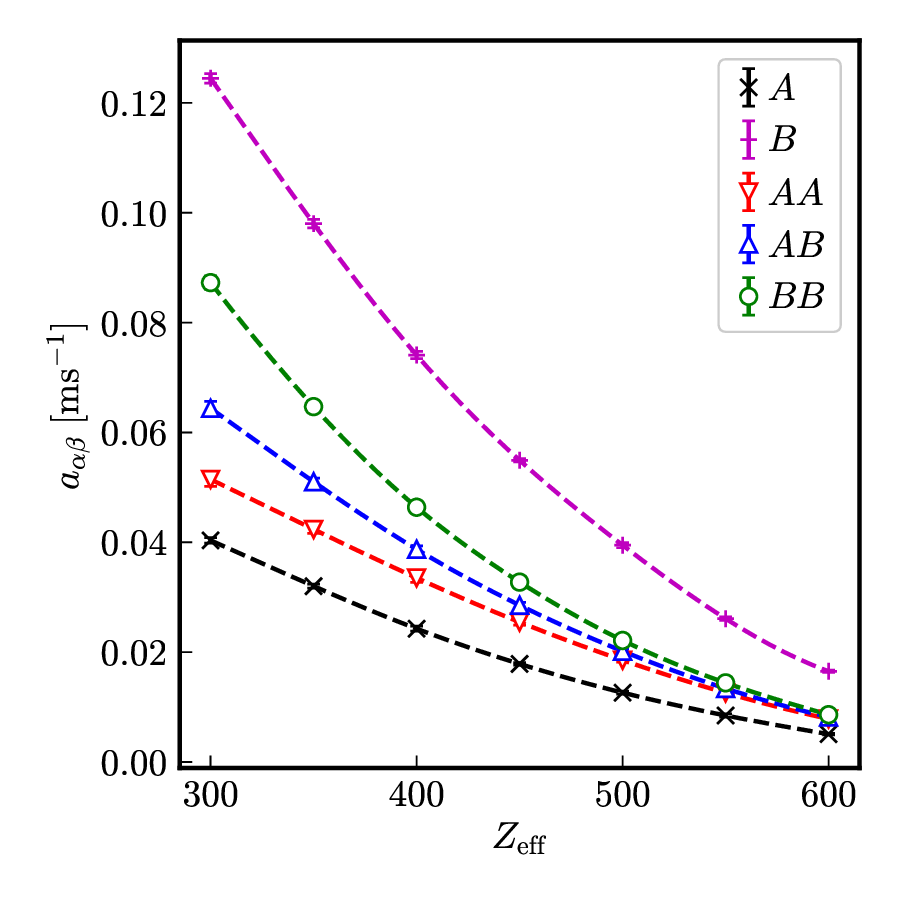}
\caption{Relaxation rates $a_{\alpha\beta}$ in a binary mixture of particles with diameters $\sigma_{\rm A}=100\,{\rm nm}$ and $\sigma_{\rm B}=33\,{\rm nm}$ in dependence on the number of effective charges $Z_{\rm eff}=Z_{\rm eff,A}=Z_{\rm eff,B}$ at identical number densities  $^1\rho_{\rm A} = \,^1\rho_{\rm B} = \SI{5e17}{\per\cubic\meter}$. For comparison, the corresponding relaxation rates for one-component systems consisting of particles with $\sigma_{\rm A}=100\,{\rm nm}$ ($\times$) and $\sigma_{\rm B}=33\,{\rm nm}$ (+) are displayed as well. The dashed lines are cubic splines as a guide to the eye.}
\label{fig:alpha_zeff_mix}
\end{figure}

The relaxation rates $a_{\alpha\beta}$ in binary mixtures of particles with $\sigma_{\rm A} = \SI{100}{\nano\meter}$ and $\sigma_{\rm B} = \SI{33}{\nano\meter}$ at the same number densities $^1\rho_{\rm A} = \,^1\rho_{\rm B} = \SI{5e17}{\per\cubic\meter}$ in dependence on the number of effective charges $Z_{\rm eff} = Z_{\rm eff, A} = Z_{\rm eff, B}$ is displayed in Fig. \ref{fig:alpha_zeff_mix}. For comparison, those of effective
one-component systems consisting of particles with diameters $\sigma_{\rm A} = \SI{100}{\nano\meter}$ and $\sigma_{\rm B} = \SI{33}{\nano\meter}$, respectively, are depicted as well.

In analogy to self-diffusion, also the decay of intermediate scattering functions is influenced by the presence of a differently sized species: The relaxation rates of the correlation between the larger particles are enhanced in the presence of more mobile smaller particles and vice versa. As expected, the relaxation rates of correlations between different species $a_{\rm AB}$ are in between
the relaxation rates $a_{\rm AA}$ and $a_{\rm BB}$ of correlations between identical particles. With increasing number of effective charges $Z_{\rm eff} = Z_{\rm eff, A} = Z_{\rm eff, B}$, a nonlinear decay of relaxation rates is observed, approaching a common limit in the case of very strong electrostatic interactions.

\begin{figure}
\centering
\includegraphics[width=1.0\columnwidth]{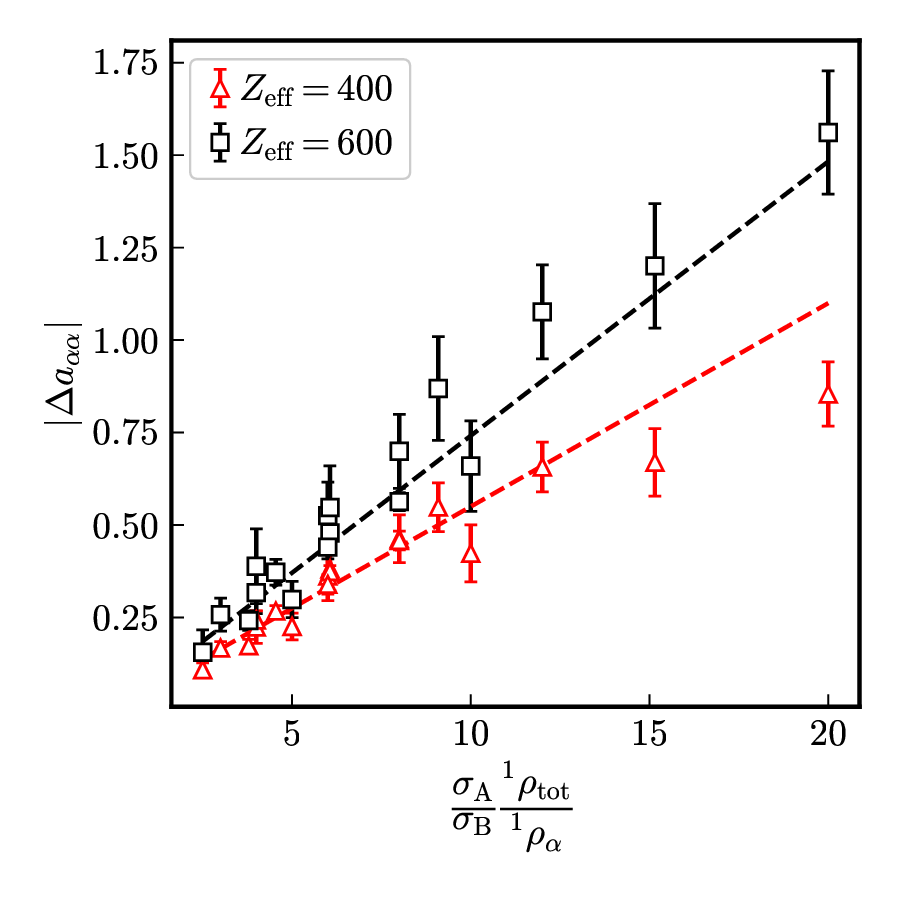}
\caption{Dependence of the modulus $\vert \Delta a_{\alpha\alpha}\vert$ of excess relaxation rates on the ratio $(\sigma_{\rm A}\,^1\rho_{\rm tot})/(\sigma_{\rm B}\,^1\rho_{\alpha})$. The dashed lines are linear fits indicating deviations from linearity at large ratios $(\sigma_{\rm A}\,^1\rho_{\rm tot})/(\sigma_{\rm B}\,^1\rho_{\alpha})$.}
\label{fig:acceleration_alpha}
\end{figure}

Defining an excess relaxation rate as
\begin{align}
\Delta a_{\alpha\alpha} &= \dfrac{a_{\alpha\alpha}-a^*_{\alpha\alpha}}{a^*_{\alpha\alpha}}
\end{align}
with $a_{\alpha\alpha}^*$ denoting the relaxation rate in a one-component system and $a_{\alpha\alpha}$ that in a binary mixture at identical total number densities. 
Analogously to self-diffusion, the modulus of excess relaxation rates of correlations between identical particles depend in first approximation linearly on the ratio $(\sigma_{\rm A}\,^1\rho_{\rm tot}) / (\sigma_{\rm B}\,^1\rho_{\alpha})$ for sufficiently strong electrostatic interaction as visible in Fig. \ref{fig:acceleration_alpha}.

\begin{figure}
\centering
\includegraphics[width=1.0\columnwidth]{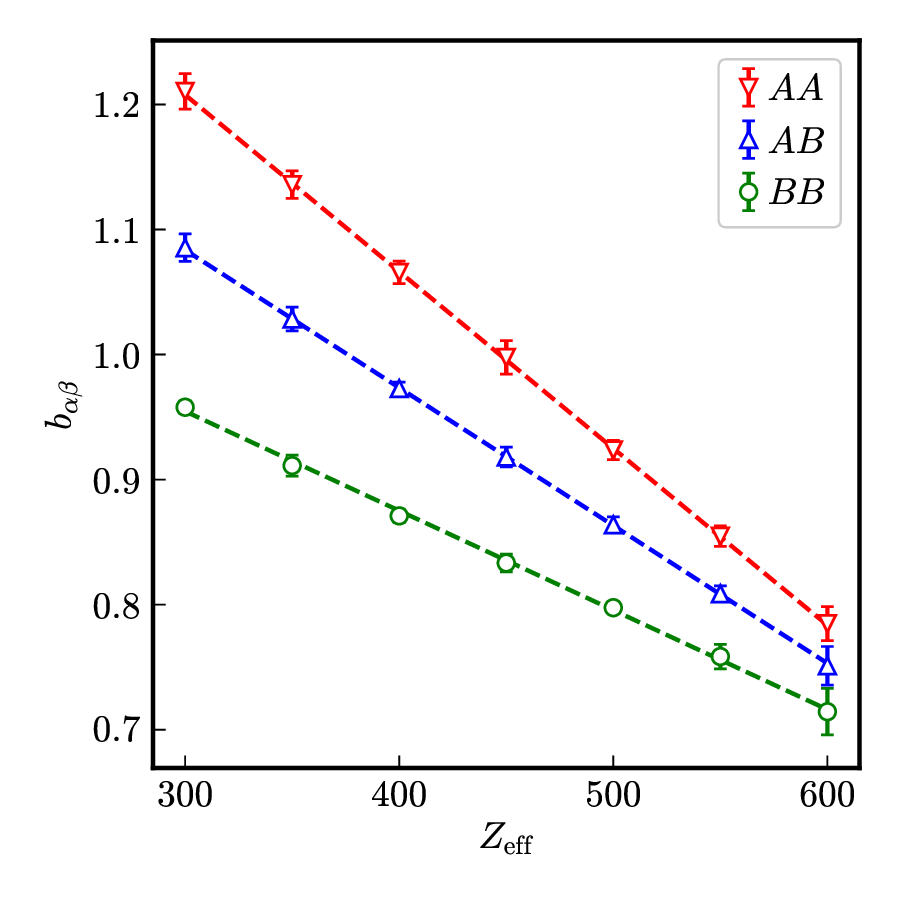}
\caption{Stretching exponents $b_{\alpha\beta}$ of a binary mixture of differently sized but identically charged colloidal macroions in dependence on their number of effective charges $Z_{\rm eff}$. The large species' diameter is $\sigma_{\rm A} = \SI{100}{\nano\meter}$ and that of the small is $\sigma_{\rm B} = \SI{33}{\nano\meter}$ at identical number densities $^1\rho_{\rm A} = \,^1\rho_{\rm B} = \SI{5e17}{\per\cubic\meter}$. The dashed lines are linear fits.}
\label{fig:beta_mix}
\end{figure}

Opposite to one-component systems, the stretching exponents $b_{\alpha\beta}$ differ in binary mixtures of equally charged particles with different sizes as exemplarily shown in Fig. \ref{fig:beta_mix} for a mixture
of particles with $\sigma_{\rm A}=100\,{\rm nm}$ and $\sigma_{\rm B}=33\,{\rm nm}$ at identical number densities $^1\rho_{\rm A}=\,^1\rho_{\rm B}=\SI{5e17}{\per\cubic\meter}$. The stretching exponents $b_{\rm AB}$ of the partial intermediate scattering functions $S_{\rm D,AB}^{\rm red}(Q_{\rm max},t)$ are for all investigated numbers of effective charges practically identical to one-component systems consisting either of the larger species $\rm A$ or the smaller species B at the same total number density. In presence of a smaller species B, the stretching exponent $b_{\rm AA}$ is larger than $b_{\rm AB}$ indicating a more compressed relaxation. The stretching exponent $b_{\rm BB}$ of the
intermediate scattering function $S_{\rm BB}(Q_{\rm max},t)$ is reduced by the presence of a larger species A indicating a more stretched relaxation. The stretching exponents $b_{\alpha\beta}$ for all investigated
mixtures are compiled in Tables \ref{tab:alpha_beta_mix_1}, \ref{tab:alpha_beta_mix_2}, and \ref{tab:alpha_beta_mix_4}.

With increasing number of effective charges and thus electrostatic repulsion, a linear decrease of all stretching exponents $b_{\alpha\beta}$ is observed, where the differences $b_{\rm AA}-b_{\rm BB}$ also decrease
with $Z_{\rm eff}$.

\begin{figure}
\centering
\includegraphics[width=1.0\columnwidth]{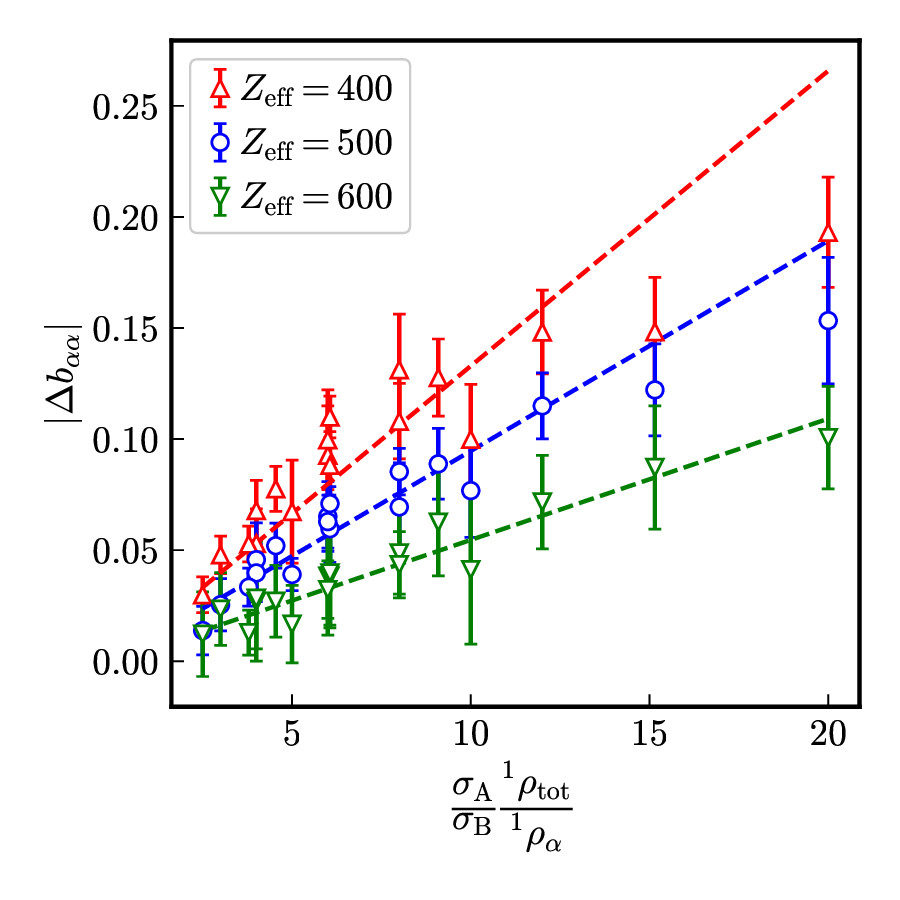}
\caption{Modulus $\left\vert \Delta b_{\alpha\alpha}\right\vert$ of the excess stretching exponent in binary mixtures of differently sized but identically charged particles in dependence on the ratio $(\sigma_{\rm A}\,^1\rho_{\rm tot}) / (\sigma_{\rm B}\,^1\rho_\alpha)$. The dashed lines are linear fits to the data.}
\label{fig:acceleration_beta}
\end{figure}

Let us analogously to the excess relaxation rate $\Delta a_{\alpha\alpha}$ define an excess stretching exponent
\begin{align}
\Delta b_{\alpha\alpha} &= b_{\alpha\alpha} - b_{\alpha}^*
\end{align}
with $b^*_\alpha$ indicating the stretching exponent in a one-component system with the same total number density.
For sufficiently strong electrostatic interaction, similar as for the reduced excess long-time self-diffusion coefficient, a nearly linear dependence on the ratio $(\sigma_{\rm A}\,^1\rho_{\rm tot}) / (\sigma_{\rm B}\,^1\rho_\alpha)$ (Fig. \ref{fig:acceleration_beta}) is observed where the slope decreases with the number of effective charges $Z_{\rm eff}$.

\subsection{Comparison of freezing criteria}

The Hansen-Verlet criterion \cite{hansen:1969} uses the mean structure factor as freezing criterion. Since in our case all partial structure factors are identical, also the mean structure factor $\Braket{S_{\alpha\beta}(Q)}_{\alpha\beta}$ is identical to the partial ones. We found that with a number of effective charges larger than $Z_{\rm eff} \gtrsim 450$ the Hansen-Verlet criterion $\max\left[\Braket{S_{\alpha\beta}(Q)}_{\alpha\beta}\right]\gtrsim 2.85$ is exceeded, predicting a frozen state for the entire binary mixture (Table \ref{tab:hansen_verlet}).

Löwen \etal \cite{Loewen:1993} proposed a reduced long-time self-diffusion coefficient $D_{\rm S,\alpha}^{\rm (L)}/D_{0,\alpha}\lesssim 0.098$ as a dynamic freezing criterion for species $\alpha$. For an effective one-component system with $Z_{\rm eff}=450$, opposite to the Hansen-Verlet criterion, the dynamic freezing criterion predicts a non-frozen state. Due to the reduced long-time self-diffusion coefficient of particles with $\sigma_{\rm B}=50\,{\rm nm}$ in presence of particles with $\sigma_{\rm A}=100\,{\rm nm}$ at identical number densities
$^1\rho_{\rm A}=\,^1\rho_{\rm B}$, this dynamic freezing criterion predicts a selectively frozen state of the more mobile subsystem B and a non-frozen state of the less mobile species A in a binary mixture. 

Both for one-component systems and binary mixtures, the long-time limits of reduced partial intermediate scattering functions $S_{\alpha\beta}^{\rm red}(Q,t\to\infty)$ decay to zero indicating a liquid-like state for all here investigated systems. At a total number density of $^1\rho_{\rm tot}=\SI{1e18}{\per\cubic\meter}$, systems with $Z_{\rm eff}\gtrsim 450$ effective charges do not melt starting from a bcc-structure. Hence, liquid-like ordered
systems with $Z_{\rm eff}\gtrsim 450$ are metastable, supercooled liquids. Also for these systems, the Debye-Waller factors vanish, indicating a non-arrested state with still complete structural relaxation.

\begin{table}[h]
\centering
\caption{Maximum $\max\left[\Braket{S_{\alpha\beta}(Q)}_{\alpha\beta}\right]$ of the mean static structure factor in dependence on the number of effective charges $Z_{\rm eff}$ of identically charged particles with diameters $\sigma_{\rm A} = \SI{100}{\nano\meter}$ and $\sigma_{\rm B} = \SI{50}{\nano\meter}$ at identical number densities $^1\rho_{\rm A} = \,^1\rho_{\rm B} = \SI{5e17}{\per\cubic\meter}$.}
\label{tab:hansen_verlet}
\begin{tabular}{llr}
\hline\hline
$Z_{\rm eff}$ & & $\max\left[\Braket{S_{\alpha\beta}(Q)}_{\alpha\beta}\right]$\\
\hline
300 & & $2.107 (23)$\\
350 & & $2.363 (24)$\\
400 & & $2.65 (4)$\\
450 & & $2.885 (25)$\\
500 & & $3.14 (4)$\\
550 & & $3.38 (4)$\\
600 & & $3.57 (5)$\\
\hline\hline
\end{tabular}
\end{table}

\section{Conclusion}

In this paper, we investigate structure and dynamics of binary Yukawa systems. Preserving the strength of electrostatical interaction, i.e., the number of effective charges,
we study the influence of different mobilities quantified by Stokes-Einstein short-time diffusion coefficients of differently sized particles. Due to practically identical interactions
between all species, also the partial correlation functions are identical. Mediated by the electrostatic interaction, however, coupling effects both in long-time self-diffusion and collective
diffusion are observed.

The long-time self-diffusion coefficients of a less mobile species are enhanced by the presence of a more mobile, smaller species and vice versa. A dynamic coupling is also visible in
collective dynamics in the relaxation rates as well as the functional form of the correlation decay in partial intermediate scattering functions calculated from time-dependent, partial
correlation functions. These coupling effects are quantified by respective excess quantities using pure systems with the same interactions and total number densities as reference systems. 
For sufficiently strong electrostatic interactions, in first approximation a linear dependence of excess quantities on the ratio $(\sigma_{\rm A}\,^1\rho_{\rm tot}) / (\sigma_{\rm B}\,^1\rho_\alpha)$ 
is identified. 

Furthermore, we compare different freezing criteria for metastable liquid-like structures. Despite for systems with $Z_{\rm eff}\gtrsim 450$ effective charges at a total number density 
of $^1\rho=\SI{1e18}{\per\cubic\meter}$ the static Hansen-Verlet criterion indicates a completely frozen state and the dynamic Löwen criterion a selectively frozen state, all intermediate scattering functions asymptotically decay to zero in the long-time
limit. Hence, a structural arrest is not observed in the here investigated systems.

Including hydrodynamic interactions is a remaining task especially at higher volume fractions where a glass transition is expected. Using the available partial static structure factors as an input, multi-component
mode coupling theory is capable to predict time-dependent processes such as self- and collective diffusion. The comparison of the here provided simulation data with multi-component mode coupling theory
is a promising test for the latter approach.

\begin{acknowledgments}
D.W. gratefully acknowledges financial support by "Stiftung Stipendien-Fonds des Verbandes der Chemischen Industrie" within their Ph.D. scholarship program.
\end{acknowledgments}

\section*{Author Declarations}

\subsection*{Conflict of Interest}
The authors have no conflict to disclose.

\subsection*{Author Contributions}
\textbf{Daniel Weidig}: Conceptualization (equal); Funding Acquisition (minor); Investigation (equal); Software (equal); Validation (equal); Visualization (equal); Writing - original draft; Writing - Editing and Revision (equal).
\textbf{Joachim Wagner}: Conceptualization (equal); Funding Acquisition (major); Investigation (equal); Software (equal); Supervision; Validation (equal); Visualization (equal); Writing - Editing and Revision (equal).

\subsection*{Data Availability}
The data that support the findings of this study are available from the corresponding author upon reasonable request.

%

\onecolumngrid

\appendix
\section{Numerical data}

\begin{table}[h]
\renewcommand{\arraystretch}{1.3}
\setlength{\tabcolsep}{4.9pt}
\centering
\caption{Reduced long-time self-diffusion coefficient $D_\mathrm{S}^{\mathrm{(L), red}}$ of one-component Yukawa-suspensions in dependence on the particle diameter $\sigma$ and number of effective charges $Z_\mathrm{eff}$ at number density $^1\rho= \SI{1e18}{\per\cubic\meter}$.}
\label{tab:dsl_pure}
\begin{tabular}{llllll}
\hline
\hline
$Z_\mathrm{eff}$ & $\sigma = 100\,\mathrm{nm}$ & $\sigma = 50\,\mathrm{nm}$ & $\sigma = 33\,\mathrm{nm}$ & $\sigma = 25\,\mathrm{nm}$ & $\sigma = 20\,\mathrm{nm}$\\
\hline
300 & 0.2771 (24) & 0.2765 (22) & 0.2757 (25) & 0.2749 (21) & 0.2761 (22)\\
350 & 0.2193 (26) & 0.2172 (20) & 0.2171 (17) & 0.2176 (18) & 0.2175 (18)\\
400 & 0.1689 (23) & 0.1687 (20) & 0.1686 (22) & 0.1678 (19) & 0.1680 (15)\\
450 & 0.1308 (21) & 0.1277 (15) & 0.1272 (15) & 0.1264 (12) & 0.1267 (12)\\
500 & 0.0907 (28) & 0.0918 (18) & 0.0919 (17) & 0.0919 (22) & 0.0909 (22)\\
550 & 0.0638 (20) & 0.0645 (23) & 0.0650 (18) & 0.0639 (16) & 0.0640 (27)\\
600 & 0.0432 (14) & 0.0435 (20) & 0.0428 (11) & 0.0422 (17) & 0.0415 (20)\\
\hline
\hline
\end{tabular}
\end{table}

\begin{table}[h]
\renewcommand{\arraystretch}{1.3}
\setlength{\tabcolsep}{4.9pt}
\centering
\caption{Reduced long-time self-diffusion coefficients $D_{\mathrm{S},\alpha}^{\mathrm{(L), red}}$ of binary colloidal suspensions in dependence on their number of effective charges $Z_\mathrm{eff}$, their diameter ratio $\sigma_\mathrm{A}/\sigma_\mathrm{B}$ and their number density ratio $^1\rho_\mathrm{B} / \,^1\rho_\mathrm{A}$. The larger species $A$ has the diameter $\sigma_\mathrm{A} = 100\,\mathrm{nm}$ and total number density is $^1\rho_{\rm tot} = \SI{1e18}{\per\cubic\meter}$.}
\label{tab:dsl_data}
\begin{tabular}{clllllllll}
\hline\hline
\multirow{2}{*}{$^1\rho_\mathrm{B}/^1\rho_\mathrm{A}$} & \multirow{2}{*}{$Z_\mathrm{eff}$} & \multicolumn{2}{c}{$\sigma_\mathrm{A}/\sigma_\mathrm{B} = 2$} & \multicolumn{2}{c}{$\sigma_\mathrm{A}/\sigma_\mathrm{B} = 3$} & \multicolumn{2}{c}{$\sigma_\mathrm{A}/\sigma_\mathrm{B} = 4$} & \multicolumn{2}{c}{$\sigma_\mathrm{A}/\sigma_\mathrm{B} = 5$}\\
 & & $D_{\mathrm{S,A}}^{\mathrm{(L), red}}$ & $D_{\mathrm{S,B}}^{\mathrm{(L), red}}$ & $D_{\mathrm{S,A}}^{\mathrm{(L), red}}$ & $D_{\mathrm{S,B}}^{\mathrm{(L), red}}$ & $D_{\mathrm{S,A}}^{\mathrm{(L), red}}$ & $D_{\mathrm{S,B}}^{\mathrm{(L), red}}$ & $D_{\mathrm{S,A}}^{\mathrm{(L), red}}$ & $D_{\mathrm{S,B}}^{\mathrm{(L), red}}$\\
\hline
\multirow{7}{*}{1} &  300 &  0.338 (6) & 0.216 (3) &  0.380 (8) & 0.182 (3) &  0.397 (5) & 0.1633 (18) &  0.419 (5) & 0.1485 (20)\\
  &  350 &  0.273 (5) & 0.1665 (22) &  0.303 (7) & 0.137 (4) &  0.324 (6) & 0.1214 (20) &  0.342 (5) & 0.1100 (18)\\
  &  400 &  0.213 (4) & 0.126 (3) &  0.240 (4) & 0.1035 (25) &  0.256 (5) & 0.0886 (20) &  0.2683 (20) & 0.0793 (15)\\
  &  450 &  0.167 (4) & 0.0952 (19) &  0.185 (4) & 0.0749 (22) &  0.195 (4) & 0.0641 (16) &  0.209 (5) & 0.0569 (15)\\
  &  500 &  0.119 (4) & 0.0658 (21) & 0.133 (4) & 0.0522 (16) & 0.144 (4) & 0.0445 (12) & 0.1511 (29) & 0.0382 (14)\\
  &  550 &  0.0827 (25) & 0.0446 (12) & 0.0948 (27) & 0.0358 (10) & 0.101 (5) & 0.0301 (18) & 0.1067 (22) & 0.0257 (8)\\
  &  600 &  0.0554 (29) & 0.0298 (17) & 0.0640 (27) & 0.0232 (12) & 0.0681 (28) & 0.0193 (10) &  0.0700 (29) & 0.0161 (7)\\
\hline
\multirow{7}{*}{2} &  300 & 0.365 (6) & 0.233 (4) & 0.417 (7) & 0.209 (4) & 0.451 (7) & 0.1941 (23) & 0.478 (8) & 0.183 (3)\\
  &  350 & 0.296 (6) & 0.182 (4) & 0.343 (6) & 0.1615 (23) & 0.372 (7) & 0.1468 (25) & 0.399 (6) & 0.1367 (21)\\
  &  400 & 0.234 (6) & 0.139 (3) & 0.273 (5) & 0.1203 (27) & 0.298 (3) & 0.1082 (22) & 0.325 (6) & 0.1018 (18)\\
  &  450 & 0.181 (5) & 0.1040 (18) & 0.214 (5) & 0.0895 (26) & 0.231 (3) & 0.0789 (13) & 0.253 (7) & 0.0729 (13)\\
  &  500 & 0.130 (4) & 0.0727 (25) & 0.154 (5) & 0.0626 (15) & 0.174 (6) & 0.0554 (15) & 0.188 (7) & 0.0504 (19)\\
  &  550 & 0.092 (4) & 0.0508 (16) & 0.112 (4) & 0.0433 (17) & 0.125 (6) & 0.0380 (13) & 0.135 (5) & 0.0337 (10)\\
  &  600 & 0.063 (4) & 0.0337 (18) & 0.074 (4) & 0.0278 (10) & 0.083 (3) & 0.0237 (12) & 0.090 (4) & 0.0215 (5)\\
\hline
\multirow{7}{*}{4} &  300 & 0.385 (10) & 0.250 (4) & 0.458 (9) & 0.2341 (22) & 0.492 (10) & 0.2235 (25) & 0.537 (12) & 0.2168 (26)\\
  &  350 & 0.314 (8) & 0.195 (4) & 0.379 (7) & 0.1814 (21) & 0.417 (7) & 0.1726 (23) & 0.454 (8) & 0.1658 (20)\\
  &  400 & 0.250 (7) & 0.150 (3) & 0.305 (9) & 0.1386 (23) & 0.340 (7) & 0.1290 (21) & 0.373 (11) & 0.1231 (21)\\
  &  450 & 0.193 (5) & 0.1127 (16) & 0.236 (8) & 0.1022 (25) & 0.267 (10) & 0.0946 (17) & 0.297 (7) & 0.0909 (20)\\
  &  500 & 0.142 (4) & 0.0802 (21) & 0.178 (6) & 0.0732 (18) & 0.202 (6) & 0.0674 (18) & 0.225 (6) & 0.0633 (17)\\
  &  550 & 0.101 (5) & 0.0563 (22) & 0.129 (6) & 0.0495 (13) & 0.148 (7) & 0.0460 (18) & 0.162 (6) & 0.0429 (15)\\
  &  600 & 0.067 (4) & 0.0361 (15) & 0.085 (6) & 0.0320 (21) & 0.098 (5) & 0.0292 (15) & 0.110 (6) & 0.0274 (15)\\
\hline\hline
\end{tabular}
\end{table}

\begin{table}[h]
\renewcommand{\arraystretch}{1.1}
\setlength{\tabcolsep}{5pt}
\centering
\caption{Exponential decay constant $a$ and stretching exponent $b$ of a virtually binary mixture in dependence on the particle diameter $\sigma$ and its number of effective charges $Z_{\rm eff}$ at identical number densities $^1\rho_{\rm A} = \,^1\rho_{\rm B} = \SI{5e17}{\per\cubic\meter}$.}
\label{tab:alpha_beta_pure}
\begin{tabular}{cllllllll}
\hline\hline
\multirow{2}{*}{$Z_\mathrm{eff}$} & \multicolumn{2}{c}{$\sigma = \SI{100}{\nano\meter}$} & \multicolumn{2}{c}{$\sigma = \SI{50}{\nano\meter}$} & \multicolumn{2}{c}{$\sigma = \SI{33}{\nano\meter}$} & \multicolumn{2}{c}{$\sigma = \SI{25}{\nano\meter}$}\\
 & $a$ / \si{\per\milli\second} & $b$ & $a$ / \si{\per\milli\second} & $b$ & $a$ / \si{\per\milli\second} & $b$ & $a$ / \si{\per\milli\second} & $b$\\
\hline
300 & 0.0404 (5) & 1.108 (10) & 0.0822 (11) & 1.115 (7) & 0.1245 (9) & 1.112 (6) & 0.1635 (18) & 1.112 (5)\\
350 & 0.0320 (4) & 1.043 (6) & 0.0642 (6) & 1.043 (4) & 0.0980 (8) & 1.047 (4) & 0.1284 (10) & 1.044 (4)\\
400 & 0.0243 (5) & 0.978 (7) & 0.0489 (4) & 0.9791 (26) & 0.0741 (7) & 0.981 (5) & 0.0988 (29) & 0.978 (18)\\
450 & 0.01786 (23) & 0.920 (5) & 0.0362 (6) & 0.920 (7) & 0.0549 (4) & 0.920 (4) & 0.0729 (6) & 0.923 (22)\\
500 & 0.01262 (28) & 0.864 (8) & 0.0255 (5) & 0.862 (4) & 0.0395 (5) & 0.869 (4) & 0.0518 (4) & 0.867 (4)\\
550 & 0.0085 (3) & 0.807 (14) & 0.0173 (5) & 0.811 (7) & 0.0261 (4) & 0.810 (5) & 0.0348 (4) & 0.8095 (27)\\
600 & 0.00508 (15) & 0.746 (10) & 0.0109 (4) & 0.756 (8) & 0.01651 (21) & 0.754 (5) & 0.0216 (5) & 0.754 (5)\\
\hline\hline
\end{tabular}
\end{table}

\begin{turnpage}
\begin{table}[h]
\renewcommand{\arraystretch}{1.1}
\setlength{\tabcolsep}{3pt}
\centering
\caption{Exponential decay constant $a$ and stretching exponent $b$ of binary Yukawa-suspensions of identical charged but differently sized colloidal particles in dependence on the number of effective charges $Z_{\rm eff}$, the ratio of the species diameters $\sigma_{\rm A}/\sigma_{\rm B}$ for a ratio of the number densities $^1\rho_{\rm B}/\,^1\rho_{\rm A} = 1$. The total number density for all suspensions is identical $^1\rho_{\rm tot} = \SI{1e18}{\per\cubic\meter}$, the diameter of species $A$ is constant $\sigma_{\rm A} = \SI{100}{\nano\meter}$.}
\label{tab:alpha_beta_mix_1}
\begin{tabular}{ccclllllllll}
\hline\hline
\multirow{2}{*}{$^1\rho_\mathrm{B}/\,^1\rho_\mathrm{A}$} & \multirow{2}{*}{$Z_\mathrm{eff}$} &  & \multicolumn{3}{c}{$\sigma_\mathrm{A}/\sigma_\mathrm{B} = 2$} & \multicolumn{3}{c}{$\sigma_\mathrm{A}/\sigma_\mathrm{B} = 3$} & \multicolumn{3}{c}{$\sigma_\mathrm{A}/\sigma_\mathrm{B} = 4$}\\
 & & $\alpha\beta$ & AA & AB & BB & AA & AB & BB & AA & AB & BB \\
\hline
\multirow{14}{*}{1} & \multirow{2}{*}{300} & $a_{\alpha\beta}$ & 0.0472 (13) & 0.0554 (8) & 0.0662 (8) & 0.0515 (13) & 0.0644 (13) & 0.0873 (13) & 0.0535 (12) & 0.0707 (14) & 0.1046 (21)\\
 & & $b_{\alpha\beta}$ & 1.177 (18) & 1.094 (11) & 1.013 (13) & 1.211 (15) & 1.085 (11) & 0.958 (6) & 1.224 (12) & 1.076 (6) & 0.918 (8)\\
 & \multirow{2}{*}{350} & $a_{\alpha\beta}$ & 0.0385 (7) & 0.0433 (5) & 0.0502 (9) & 0.0423 (8) & 0.0510 (7) & 0.0647 (9) & 0.0441 (9) & 0.0555 (7) & 0.0752 (12)\\
 & & $b_{\alpha\beta}$ & 1.101 (10) & 1.032 (6) & 0.960 (9) & 1.136 (11) & 1.028 (10) & 0.911 (9) & 1.153 (13) & 1.023 (6) & 0.879 (10)\\
 & \multirow{2}{*}{400} & $a_{\alpha\beta}$ & 0.0297 (6) & 0.0332 (5) & 0.0367 (10) & 0.0335 (8) & 0.0387 (7) & 0.0464 (9) & 0.0355 (10) & 0.0423 (8) & 0.0536 (10)\\
 & & $b_{\alpha\beta}$ & 1.030 (10) & 0.970 (7) & 0.911 (12) & 1.066 (9) & 0.973 (6) & 0.871 (6) & 1.086 (11) & 0.971 (5) & 0.847 (8)\\
 & \multirow{2}{*}{450} & $a_{\alpha\beta}$ & 0.0225 (6) & 0.0243 (10) & 0.0264 (5) & 0.0255 (6) & 0.0285 (6) & 0.0328 (8) & 0.0270 (7) & 0.0311 (6) & 0.0369 (6)\\
 & & $b_{\alpha\beta}$ & 0.966 (7) & 0.918 (10) & 0.867 (8) & 0.998 (14) & 0.918 (8) & 0.833 (8) & 1.007 (7) & 0.915 (9) & 0.811 (6)\\
 & \multirow{2}{*}{500} & $a_{\alpha\beta}$ & 0.0166 (5) & 0.0174 (4) & 0.0184 (4) & 0.0185 (4) & 0.0202 (4) & 0.0221 (5) & 0.0196 (6) & 0.0218 (6) & 0.0248 (6)\\
 & & $b_{\alpha\beta}$ & 0.910 (10) & 0.866 (9) & 0.823 (9) & 0.924 (8) & 0.864 (7) & 0.798 (5) & 0.933 (13) & 0.862 (11) & 0.782 (8)\\
 & \multirow{2}{*}{550} & $a_{\alpha\beta}$ & 0.0111 (4) & 0.0116 (5) & 0.0121 (5) & 0.01258 (28) & 0.01338 (28) & 0.0144 (4) & 0.0137 (3) & 0.01468 (29) & 0.0161 (5)\\
 & & $b_{\alpha\beta}$ & 0.838 (10) & 0.811 (10) & 0.780 (10) & 0.855 (9) & 0.809 (7) & 0.758 (10) & 0.865 (8) & 0.809 (8) & 0.751 (10)\\
 & \multirow{2}{*}{600} & $a_{\alpha\beta}$ & 0.0070 (4) & 0.0072 (3) & 0.0074 (4) & 0.0079 (4) & 0.0082 (5) & 0.0086 (6) & 0.00863 (26) & 0.0090 (4) & 0.0094 (4)\\
 & & $b_{\alpha\beta}$ & 0.773 (17) & 0.750 (15) & 0.728 (15) & 0.785 (14) & 0.751 (16) & 0.714 (19) & 0.795 (9) & 0.755 (9) & 0.711 (11)\\
 \hline\hline
\end{tabular}
\end{table}

\begin{table}[h]
\renewcommand{\arraystretch}{1.1}
\setlength{\tabcolsep}{3pt}
\centering
\caption{Exponential decay constant $a$ and stretching exponent $b$ of binary Yukawa-suspensions of identical charged but differently sized colloidal particles in dependence on the number of effective charges $Z_{\rm eff}$, the ratio of the species diameters $\sigma_{\rm A}/\sigma_{\rm B}$ for a ratio of the number densities $^1\rho_{\rm B}/\,^1\rho_{\rm A} = 2$. The total number density for all suspensions is identical $^1\rho_{\rm tot} = \SI{1e18}{\per\cubic\meter}$, the diameter of species $A$ is constant $\sigma_{\rm A} = \SI{100}{\nano\meter}$.}
\label{tab:alpha_beta_mix_2}
\begin{tabular}{ccclllllllll}
\hline\hline
\multirow{2}{*}{$^1\rho_\mathrm{B}/\,^1\rho_\mathrm{A}$} & \multirow{2}{*}{$Z_\mathrm{eff}$} &  & \multicolumn{3}{c}{$\sigma_\mathrm{A}/\sigma_\mathrm{B} = 2$} & \multicolumn{3}{c}{$\sigma_\mathrm{A}/\sigma_\mathrm{B} = 3$} & \multicolumn{3}{c}{$\sigma_\mathrm{A}/\sigma_\mathrm{B} = 4$}\\
 & & $\alpha\beta$ & AA & AB & BB & AA & AB & BB & AA & AB & BB \\
\hline
\multirow{14}{*}{2} & \multirow{2}{*}{300} & $a_{\alpha\beta}$ & 0.0504 (18) & 0.0588 (10) & 0.0709 (13) & 0.0559 (16) & 0.0709 (14) & 0.0979 (14) & 0.0594 (11) & 0.0787 (14) & 0.1210 (16)\\
 & & $b_{\alpha\beta}$ & 1.196 (19) & 1.117 (9) & 1.038 (11) & 1.251 (22) & 1.122 (11) & 0.991 (4) & 1.268 (20) & 1.114 (11) & 0.954 (8)\\
 & \multirow{2}{*}{350} & $a_{\alpha\beta}$ & 0.0409 (7) & 0.0466 (9) & 0.0541 (11) & 0.0465 (10) & 0.0566 (11) & 0.0733 (11) & 0.0498 (12) & 0.0639 (9) & 0.0901 (8)\\
 & & $b_{\alpha\beta}$ & 1.128 (11) & 1.059 (10) & 0.981 (9) & 1.175 (17) & 1.064 (11) & 0.946 (7) & 1.201 (10) & 1.062 (8) & 0.918 (8)\\
 & \multirow{2}{*}{400} & $a_{\alpha\beta}$ & 0.0330 (10) & 0.0363 (9) & 0.0407 (6) & 0.0376 (9) & 0.0442 (8) & 0.0543 (7) & 0.0402 (9) & 0.0496 (9) & 0.0653 (8)\\
 & & $b_{\alpha\beta}$ & 1.070 (16) & 0.999 (10) & 0.931 (6) & 1.105 (12) & 1.010 (6) & 0.903 (7) & 1.126 (13) & 1.007 (7) & 0.879 (5)\\
 & \multirow{2}{*}{450} & $a_{\alpha\beta}$ & 0.0246 (10) & 0.0267 (15) & 0.0292 (11) & 0.0289 (9) & 0.0328 (7) & 0.0388 (8) & 0.0320 (10) & 0.0375 (6) & 0.0458 (8)\\
 & & $b_{\alpha\beta}$ & 0.992 (17) & 0.938 (14) & 0.882 (11) & 1.028 (13) & 0.945 (10) & 0.861 (6) & 1.056 (14) & 0.955 (4) & 0.841 (7)\\
 & \multirow{2}{*}{500} & $a_{\alpha\beta}$ & 0.0181 (6) & 0.0191 (4) & 0.0205 (5) & 0.0213 (6) & 0.0237 (7) & 0.0265 (7) & 0.0236 (4) & 0.02676 (24) & 0.0311 (6)\\
 & & $b_{\alpha\beta}$ & 0.929 (9) & 0.881 (5) & 0.837 (8) & 0.953 (9) & 0.890 (8) & 0.817 (8) & 0.979 (8) & 0.897 (7) & 0.804 (9)\\
 & \multirow{2}{*}{550} & $a_{\alpha\beta}$ & 0.0123 (4) & 0.0128 (4) & 0.0134 (5) & 0.0149 (6) & 0.0161 (6) & 0.0174 (8) & 0.0165 (5) & 0.01802 (26) & 0.0201 (5)\\
 & & $b_{\alpha\beta}$ & 0.853 (11) & 0.821 (6) & 0.786 (8) & 0.884 (11) & 0.833 (10) & 0.777 (12) & 0.899 (11) & 0.834 (5) & 0.765 (6)\\
 & \multirow{2}{*}{600} & $a_{\alpha\beta}$ & 0.00774 (24) & 0.00787 (27) & 0.00807 (25) & 0.0095 (5) & 0.0099 (5) & 0.0104 (5) & 0.0105 (4) & 0.0113 (4) & 0.0121 (5)\\
 & & $b_{\alpha\beta}$ & 0.785 (17) & 0.756 (10) & 0.733 (9) & 0.809 (15) & 0.770 (13) & 0.727 (12) & 0.818 (12) & 0.772 (7) & 0.722 (9)\\
\hline\hline
\end{tabular}
\end{table}

\begin{table}[h]
\renewcommand{\arraystretch}{1.1}
\setlength{\tabcolsep}{3pt}
\centering
\caption{Exponential decay constant $a$ and stretching exponent $b$ of binary Yukawa-suspensions of identical charged but differently sized colloidal particles in dependence on the number of effective charges $Z_{\rm eff}$, the ratio of the species diameters $\sigma_{\rm A}/\sigma_{\rm B}$ for a ratio of the number densities $^1\rho_{\rm B}/\,^1\rho_{\rm A} = 4$. The total number density for all suspensions is identical $^1\rho_{\rm tot} = \SI{1e18}{\per\cubic\meter}$, the diameter of species $A$ is constant $\sigma_{\rm A} = \SI{100}{\nano\meter}$.}
\label{tab:alpha_beta_mix_4}
\begin{tabular}{ccclllllllll}
\hline\hline
\multirow{2}{*}{$^1\rho_\mathrm{B}/\,^1\rho_\mathrm{A}$} & \multirow{2}{*}{$Z_\mathrm{eff}$} &  & \multicolumn{3}{c}{$\sigma_\mathrm{A}/\sigma_\mathrm{B} = 2$} & \multicolumn{3}{c}{$\sigma_\mathrm{A}/\sigma_\mathrm{B} = 3$} & \multicolumn{3}{c}{$\sigma_\mathrm{A}/\sigma_\mathrm{B} = 4$}\\
 & & $\alpha\beta$ & AA & AB & BB & AA & AB & BB & AA & AB & BB \\
\hline
\multirow{14}{*}{4} & \multirow{2}{*}{300} & $a_{\alpha\beta}$ & 0.0520 (27) & 0.0615 (14) & 0.0748 (12) & 0.0592 (18) & 0.0760 (18) & 0.1072 (17) & 0.0637 (28) & 0.0862 (11) & 0.1362 (11)\\
 & & $b_{\alpha\beta}$ & 1.22 (4) & 1.139 (10) & 1.057 (9) & 1.26 (29) & 1.150 (13) & 1.024 (8) & 1.31 (5) & 1.154 (9) & 0.999 (7)\\
 & \multirow{2}{*}{350} & $a_{\alpha\beta}$ & 0.0429 (14) & 0.0489 (8) & 0.0576 (8) & 0.0500 (7) & 0.0617 (8) & 0.0818 (9) & 0.0542 (15) & 0.0706 (10) & 0.1034 (9)\\
 & & $b_{\alpha\beta}$ & 1.146 (15) & 1.073 (9) & 1.000 (7) & 1.217 (28) & 1.094 (9) & 0.974 (7) & 1.244 (22) & 1.099 (9) & 0.957 (5)\\
 & \multirow{2}{*}{400} & $a_{\alpha\beta}$ & 0.0345 (13) & 0.0384 (7) & 0.0436 (7) & 0.0405 (15) & 0.0488 (6) & 0.0612 (8) & 0.0450 (13) & 0.0561 (10) & 0.0765 (14)\\
 & & $b_{\alpha\beta}$ & 1.077 (19) & 1.012 (10) & 0.949 (6) & 1.126 (19) & 1.035 (7) & 0.928 (4) & 1.171 (19) & 1.045 (8) & 0.911 (6)\\
 & \multirow{2}{*}{450} & $a_{\alpha\beta}$ & 0.0263 (10) & 0.0287 (13) & 0.0318 (9) & 0.0318 (10) & 0.0376 (7) & 0.0448 (7) & 0.0364 (11) & 0.0433 (8) & 0.0551 (11)\\
 & & $b_{\alpha\beta}$ & 1.013 (20) & 0.953 (12) & 0.897 (6) & 1.058 (17) & 0.978 (10) & 0.884 (5) & 1.098 (18) & 0.988 (8) & 0.871 (6)\\
 & \multirow{2}{*}{500} & $a_{\alpha\beta}$ & 0.0195 (8) & 0.0210 (6) & 0.0225 (6) & 0.0242 (7) & 0.0271 (7) & 0.0308 (6) & 0.0275 (13) & 0.0318 (9) & 0.0380 (7)\\
 & & $b_{\alpha\beta}$ & 0.941 (14) & 0.898 (10) & 0.849 (7) & 0.986 (14) & 0.915 (9) & 0.835 (6) & 1.017 (22) & 0.927 (10) & 0.828 (4)\\
 & \multirow{2}{*}{550} & $a_{\alpha\beta}$ & 0.0134 (6) & 0.0141 (4) & 0.0149 (4) & 0.0174 (8) & 0.0185 (6) & 0.0203 (6) & 0.0195 (7) & 0.0221 (5) & 0.0250 (6)\\
 & & $b_{\alpha\beta}$ & 0.872 (12) & 0.834 (10) & 0.798 (7) & 0.914 (21) & 0.850 (9) & 0.788 (8) & 0.935 (15) & 0.864 (6) & 0.787 (9)\\
 & \multirow{2}{*}{600} & $a_{\alpha\beta}$ & 0.0084 (4) & 0.0089 (4) & 0.0092 (4) & 0.0112 (6) & 0.0118 (4) & 0.01253 (28) & 0.0130 (5) & 0.0139 (6) & 0.0152 (8)\\
 & & $b_{\alpha\beta}$ & 0.788 (24) & 0.770 (15) & 0.744 (12) & 0.834 (18) & 0.790 (7) & 0.741 (6) & 0.847 (14) & 0.796 (12) & 0.738 (13)\\
\hline\hline
\end{tabular}
\end{table}

\end{turnpage}

\twocolumngrid

\end{document}